\documentclass[usenatbib]{mn2e}
\usepackage{epsfig}
\usepackage{amsmath}
\usepackage{amssymb}
\title[Star formation histories of group galaxies]{The Dawn of the Red:
  Star formation histories of group galaxies over the past 5 billion years}

\author[McGee et al.]{Sean L. McGee$^{1,2}$\thanks{Email: s.l.mcgee@durham.ac.uk}, Michael L. Balogh$^{1}$,  David J. Wilman$^{3}$, Richard G. Bower$^{2}$, 
\newauthor John S. Mulchaey$^{4}$, Laura C. Parker$^{5}$ and Augustus
Oemler, Jr.$^{4}$
\\
$^{1}$Department of Physics and Astronomy, University of Waterloo, Waterloo, Ontario, N2L 3G1, Canada\\
$^{2}$Department of Physics, University of Durham, Durham, UK, DH1 3LE\\
$^{3}$Max--Planck--Institut f{\" u}r extraterrestrische Physik, Giessenbachstrasse 85748 Garching Germany\\
$^{4}$Observatories of the Carnegie Institution, 813 Santa Barbara Street, Pasadena, California, USA\\
$^{5}$Department of Physics and Astronomy, McMaster University, Hamilton, Ontario L8S 4M1, Canada\\
}

\date{\today}

\def\Mtrunc{$M_{\mathrm{trunc}}$}
\def\Ttrunc{$T_{\mathrm{trunc}}$}
\def\LCDM{$\Lambda$CDM$~$}
\def\Mdoth{$h^{-1}~$M$_\odot$}
\def\Mdothspace{$h^{-1}~$M$_\odot~$}
\def\Mdot{M$_\odot$}
\def\Mdotspace{M$_\odot~$}
\def\Mpch{$h_{75}^{-1}~$Mpc$~$}
\def\hkpc{$h^{-1}~$kpc$~$}
\def\etal{et al.$~$}

\def\kmsmpc{\>{\rm km}\,{\rm s}^{-1}\,{\rm Mpc}^{-1}}
\def\galex{{\it GALEX}\ }   
\def\swarp{\textsc{swarp}\ }   
\def\sextractor{\textsc{SExtractor}\ }   
\def\arcsec{''}
\def\Mstellar{$M_\odot~$}
\def\arcmin{'}
\def\micron{\mu\mathrm{m}}

\begin{document}
\maketitle

\begin{abstract}

We examine the star formation properties of group and field galaxies
in two surveys, SDSS (at z $\sim$ 0.08) and GEEC (at z $\sim$
0.4). Using UV imaging from the \galex space telescope, along with
optical and, for GEEC, near infrared photometry, we compare the
observed spectral energy distributions to large suites of stellar
population synthesis models. This allows us to accurately determine
star formation rates and stellar masses. We find that star forming
galaxies of all environments undergo a systematic lowering of their
star formation rate between z=0.4 and z=0.08 regardless of
mass. Nonetheless, the fraction of passive galaxies is higher in
groups than the field at both redshifts.  Moreover, the difference
between the group and field grows with time and is mass-dependent, in
the sense the the difference is larger at low masses.  However, the
star formation properties of star forming galaxies, as measured by
their average specific star formation rates, are consistent within the
errors in the group and field environment at fixed redshift. The
evolution of passive fraction in groups between z=0.4 and z=0 is
consistent with a simple accretion model, in which galaxies are
environmentally affected 3 Gyrs after falling into a $\sim 10^{13}$
\Mdotspace group.  This long timescale appears to be inconsistent with the
need to transform galaxies quickly enough to ensure that star forming
galaxies appear similar in both the group and field, as observed.

\end{abstract}


\section{Introduction}

The star formation history of a galaxy is a function of, at least,
stellar mass, redshift and environment. In the local universe, a
higher fraction of low mass galaxies are actively forming stars than
more massive galaxies \citep[][B04]{kauffmann_mass,
brinchmann_sfr}. It has been known for some time that the star
formation density of the universe has decreased by at least a factor
of 10 in the last 8 or 10 Gyrs \citep{lilly96, madau96,
hopkins2004}. There is growing evidence that this reduction with time
is seen at all stellar masses \citep{gilbank_roles}. Finally, at least
in the local universe, the fraction of star forming galaxies in groups
and clusters at fixed stellar mass is lower than the general field
\citep{kauffmann_envt, Kimm+09}. Untangling why and to what extent
each of stellar mass, redshift and environment determine a galaxy's
properties is a fundamental goal of galaxy formation and evolution
research. Ultimately, we hope to uncover the physical mechanisms
responsible for each correlation.

The role of environment has been studied extensively by many groups
 \citep[for recent reviews, see][]{boselli_review,
 blanton_review}. Importantly, \citet{weinmann} showed, by separating
 galaxies based on colour and specific star formation rate (based on
 H$\alpha$), that the fraction of blue, star forming galaxies
 decreases with increasing halo mass at fixed luminosity.  More
 recently, using UV derived star formation rates and the same group
 catalogue, \citet{Kimm+09} finds that the fraction of passive
 satellites increases with halo mass. These studies were largely
 focused on the fraction of passive galaxies, rather than the actual
 star formation rates.

Surprisingly, however, there is evidence that galaxies that are
forming stars in groups have similar properties to those in the field;
it is just that the fraction of those galaxies varies. For example,
many authors \citep[eg.][]{strateva,baldry} have found that the local
galaxy distribution is bimodal in colour, having a red and a blue
peak. \citet{balogh_bimodal} finds that the peak of the red and blue
galaxies change relative heights with environment at fixed luminosity.
However, importantly, they find no significant difference in the
location of the blue peak with environment \citep[but see
][]{wilman_multi}. Recently, \citet{Peng+10} found that the
relationship between star formation rate and stellar mass was the same
in the highest and lowest density quartile of galaxies. This implies
the mechanism that transforms galaxies in dense environments must be
rapid.

Strangulation, the process in which the hot gas halo surrounding a
galaxy is stripped when it becomes a satellite in a large dark matter
halo, is often thought to act over timescales of $>$ 2 Gyrs, seemingly
in contradiction with the observational need for a rapid timescale
\citep{mccarthy}. Therefore, a process that involves the ram pressure
stripping of a galaxy's cold gas seems more viable. However, after
correcting for the finite number of member galaxies, \citet{Balogh+10}
found that the intrinsic scatter between the red fractions in
individual galaxy groups and clusters is remarkably small. When
directly compared to models for how galaxies are accreted into groups
and subsequently into clusters, this small scatter suggests that star
formation must be truncated in haloes with mass $\leq$ 10$^{13}$
\Mdoth \citep{mcgee_accretion}.  The efficiency of ram pressure
stripping of cold gas in such low mass haloes is likely to be poor.

The next step is to look at the evolution of the key observational
properties for further clues about the relevant mechanisms. It is
difficult to obtain a large collection of unbiased and well sampled
galaxy groups at high and intermediate redshift and thus the majority
of the previous work has been based in the local Universe. To combat
this, our collaboration, the Group Environment Evolution Collaboration
(GEEC), has undertaken a detailed, multi-wavelength study of galaxy
groups at intermediate redshift (0.3 $<$ z $<$ 0.55)
\citep{Wilman1}. We have shown that these are truly galaxy groups ---
rather than clusters --- as the group-sized velocity dispersions agree
with stacked weak lensing and X-ray luminosities \citep{parker,
Finoguenov+09}. We have shown that the morphology-environment
relation, using either visual or quantitative morphologies, while in
place at z=0.4, grows stronger to z=0 \citep{mcgee, wilmanS0}. In
addition, as in the local universe, the fraction of [OII] emitting
galaxies, infrared excess galaxies or blue galaxies as a function of
stellar mass is higher in the field than in the groups
\citep{Balogh_smass, wilmanIRAC, Balogh_cnoccol}.

[OII] emission can be effectively corrected to be a useful tracer of
average star formation rates for large samples \citep{moustakas,
gilbank_sdss}. However, it is not clear that these corrections are
effective for subsamples of galaxies, such as those that are affected
by dense environments \citep{yan_oii, lemaux_oii}. In addition, to
properly separate star forming from non-star forming galaxies, it is
necessary to probe low star formation rates. It is difficult to attain
this level of sensitivity with [OII], especially in low signal to
noise spectra. In this paper, we use SED-fit star formation rates,
that are driven by UV data from \galex, and stellar masses, that are
driven by $K$ band data. This allows us to probe how star formation
evolves as a function of environment and stellar mass since z=0.5.

In \textsection \ref{sec-data}, we explain the two distinct surveys
(GEEC and SDSS) used in the paper as well as the wide array of
photometric and spectroscopic data used in both. We will also detail
the new \galex (\textsection \ref{sec-galex}) and CFHT Wircam K band
(\textsection \ref{sec-kband}) observations that will be used to fit
detailed spectral energy distributions. In \textsection \ref
{sec-SED}, we explain the SED fits used in this paper including the
fitting methodology and the sample of comparison stellar populations
that are used to derive physical parameters. Finally, we simulate a
sample of galaxies that allow tests of the robustness and accuracy
with which we recover physical parameters (\textsection
\ref{sec-simlgals}). In \textsection \ref{sec-results}, we examine the
environmental dependence of the SSFR - \Mstellar diagram at both
redshift epochs. In \textsection \ref{sec-discussion} we discuss the
results and derive a plausible toy model for the truncation of star
formation in group galaxies. Finally, in \textsection
\ref{sec-conclusions}, we discuss our conclusions.  In Appendix
\ref{sec-physical}, we make direct comparisons between our method of
determining physical parameters and other methods from the literature.

Throughout this paper, we adopt a \LCDM cosmology with the parameters;
$\Omega_{\rm m} = 0.3$, $\Omega_{\Lambda}=0.7$ and $h=H_0/(100
\kmsmpc)=0.75$. Also, in this paper all magnitudes are stated within
the AB magnitude system \citep{Oke+83}.

\section{The Data}\label{sec-data}

To achieve our goal of studying the evolution of the star formation
properties of galaxies and their dependence on environment requires a
sample of galaxies at two redshift epochs and deep UV
observations. For our z=0.4 sample, we use the sample of galaxies in
the Group Environment Evolution Collaboration (GEEC) and deep \galex
observations made in Guest Observer mode. The low redshift sample of
galaxies is derived from the Sloan Digital Sky Survey (SDSS) and its
overlap with the \galex Medium Imaging Survey (MIS). Below we detail
each sample, as well as the group finding algorithms used.

\subsection{GEEC survey}

The GEEC survey was designed to provide highly complete and deep
spectroscopy in the fields of galaxy groups selected from an earlier,
sparsely sampled redshift survey -- the Canadian Network for
Observational Cosmology (CNOC2) Field Galaxy Redshift Survey. CNOC2 is
a spectroscopic and photometric survey completed with the Multi-Object
Spectroscopy (MOS) instrument at the 3.6m Canada France Hawaii
telescope (CFHT) \citep{CNOC2}. The original survey targeted galaxies
over four different patches of sky totaling about 1.5 square
degrees. Spectroscopic redshifts of 6000 galaxies were obtained with
an overall sampling rate of 48$\%$ to R$_C$=21.5. The survey used a
band limiting filter that primarily limited successful redshifts to
the range 0.1 $<$ z $<$ 0.6, however, there was no colour
pre-selection of targets.

Despite this relatively sparse sampling, \citet{CNOC2-groups} were
able to use a slightly modified friends of friends algorithm to define
a sample of $\sim$ 200 galaxy groups. The only modification to the
standard friends of friends algorithm, such as that described by
\citet{huchra}, is the requirement that the grouped galaxies are in a
large scale overdensity. However, because of the sparse sampling of
the survey, this step has no effect on the type of groups discovered.

These galaxy groups have been robustly characterised by weak lensing
\citep{parker}, mock catalogues \citep{mcgee}, X-ray imaging
\citep{Finoguenov+09} and total stellar mass \citep{Balogh_smass}. The
mass estimates from all of these approaches agree well with the
expected masses based on the average velocity dispersions of the
groups. The mock catalogue analysis also found that the fraction of
galaxies not associated with real galaxy groups was only 2.5$\%$
(ie. those not within 0.5 \Mpch of a $>$ 10$^{12}$ \Mdothspace halo).

The first step of the GEEC survey was presented in \citet{Wilman1}, in
which the collaboration obtained follow-up spectroscopy in 20 regions,
each centered on a Carlberg \etal group at 0.3 $<$ z $<$ 0.55. This
spectroscopy was obtained using the Multi-Object Spectroscopy Low
Dispersion Survey Spectrograph (LDSS2) at the 6.5m Baade telescope at
Las Companas Observatory (LCO) in Chile.  This targeting greatly
increased the sampling in these 20 groups, as well as six other
Carlberg \etal groups that were partially overlapping our targets on
the sky. Thus, the completeness rose to $\sim$ 78$\%$ to a statisical
limit of R$_c$=22. This was a half magnitude deeper than the original
CNOC2 survey.

In this paper, we use data from both the targeted GEEC groups and the
original Carlberg \etal groups, as well as making use of the full
CNOC2 area from which to draw a comparison field sample. We correct for
the incomplete sampling by adopting the weighting scheme described in
\citet{Balogh_cnoccol}. We apply weights to the galaxies based only
the completeness as a function of $R_c$ magnitude from the original
CNOC2 images and their location inside or outside the LDSS2 follow-up
regions. This weighting scheme allows a statistically complete sample
to $R_c$ = 22 in the LDSS2 regions and to $R_c$ = 21.5 outside this
region. Further 1/V$_{\mathrm{max}}$ weighting is applied to construct
a statistically complete volume limited sample.

\subsubsection{\galex Observations}\label{sec-galex}

The Galaxy Evolution Explorer (\galex) satellite is a NASA Small
Explorer Class mission launched in 2003, which provides both imaging
and grism capabilities in the ultraviolet
\citep{Martin+05,Morrissey+07}. The \galex satellite's dichroic
beamsplitter feeds both a Near Ultraviolet (NUV;
$\lambda_\mathrm{effective}$ = 2271\AA ) and a Far Ultraviolet (FUV;
$\lambda_\mathrm{effective}$ = 1528\AA) detector, thereby providing
simultaneous imaging in both bands over a very large, circular area
(1.$^{\circ}$25 diameter).  The \galex satellite has been
revolutionizing galaxy formation and evolution studies since its
launch by providing such wide field imaging in the ultraviolet. This
is a particularly important waveband to enable a characterization of
the hot, massive and young stars that are indicative of recent star
formation. Although, as we discuss, corrections must be made for both
dust extinction and contributions from old stellar populations before
robust star formation rates can be obtained.

We were awarded 9 orbits of \galex observation time ($\sim$ 13.5 ks
total) in Cycle 1 (PI: M. Balogh, ID: 037). This was intended to allow
three orbits of observation time for each of three of the four CNOC2
patches. However, early in the \galex mission, intermittent problems
with the FUV detector caused some observations to be obtained with
only the NUV images passing image quality tests. Thus, further
observations were undertaken to insure that the full compliment of FUV
observing time was achieved. For this reason, the exposure time of the
NUV images vary from 3 orbits per field (the 2 hour patch) to 6 orbits
per field (the 14 hour patch).  In Table \ref{table-geecgalexobs}, we
detail the \galex data of the three CNOC2 patches. Because of the
L-shaped geometry of the original CNOC2 fields, the circular \galex
pointings do not cover the full patch. In this paper we restrict our
analysis to galaxies within 0.$^{\circ}$6 of the center of the \galex
pointings.

\begin{table}
\begin{center}
\begin{tabular}{| c | c | c | c | c | }
\hline
Patch & Central RA & Central DEC  & NUV time  & FUV time \\
 & (J2000) & (J2000)  &  (seconds) & (seconds) \\
\hline
 2h  &  2:26:03.9 & 00:21:34 & 5843 & 5676  \\
 14h   &  14:49:38.1 & 09:10:58 & 11739 & 5115  \\
 21h   & 21:51:20.3 & -05:31:27 & 8546 & 4682 \\
\hline
\end{tabular}
\end{center}
\caption[\galex observations of the GEEC survey]{Details of the \galex
observations of the GEEC survey. Listed are the original CNOC2 patch
name, the central right ascension (RA) and declination (DEC) of the
\galex pointings, as well as the combined FUV and NUV exposure time.}
\label{table-geecgalexobs}

\end{table}

\begin{figure*}
\includegraphics[width=\textwidth]{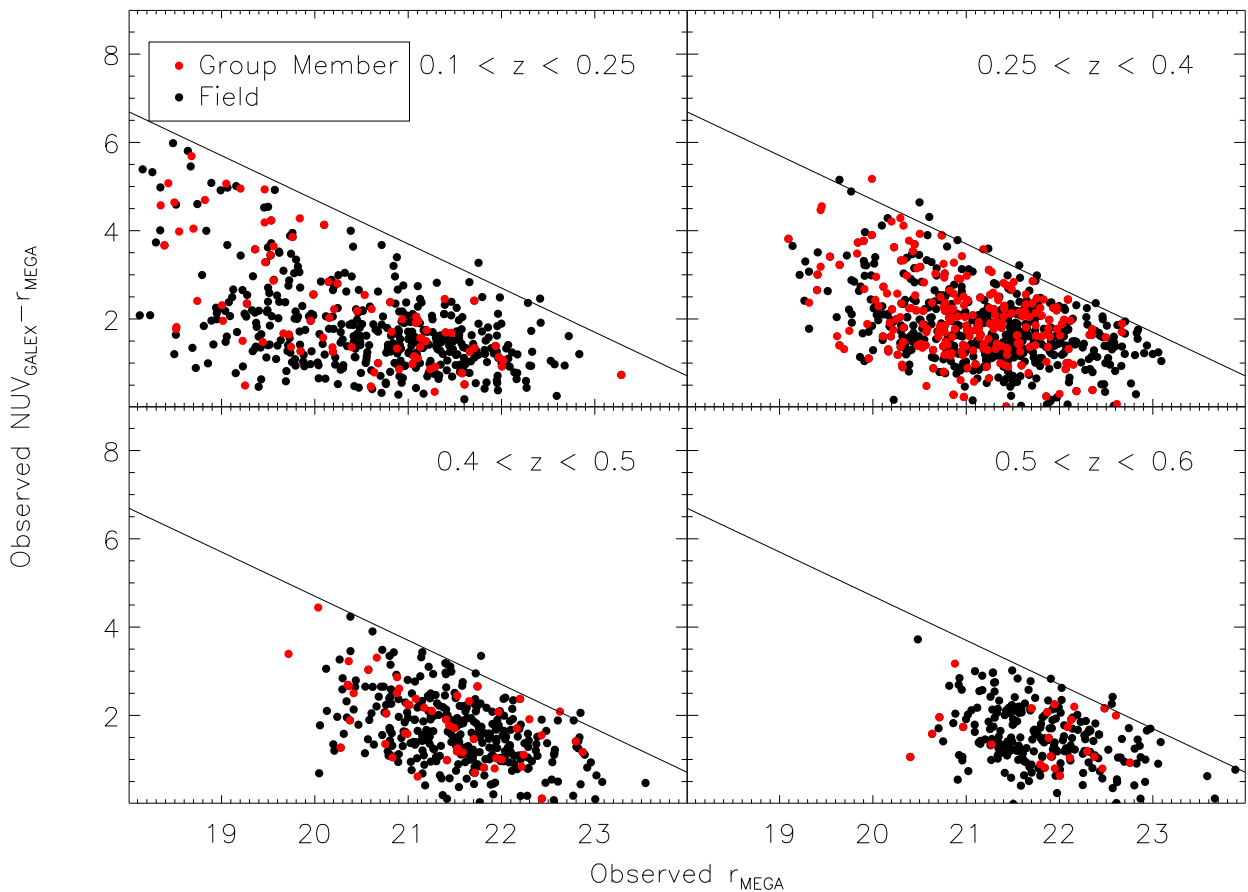}
\caption[Observed colour magnitude in the \galex NUV and CFHT Megacam
 r bands in four redshift bins]{Observed colour-magnitude diagram in the
 \galex NUV and CFHT Megacam r bands in four redshift bins within the
 GEEC survey. The black line is at NUV=24.7, which illustrates the
 limit of the faintest sources detected in the \galex images. The red
 points are group members while the black points are field galaxies. }
\label{fig-observedCMD}
\end{figure*}

We use the standard \galex image reduction and calibrations obtained
from the 2nd data release \galex pipeline, as outlined in
\citet{Morrissey+07}. To provide a demonstration of the depth of our
observations we present Figure \ref{fig-observedCMD}, which shows the
observed NUV-r color as a function of observed r magnitude in four
different redshift bins. In this Figure, we have matched sources that
are individually detected on the NUV images to the nearest GEEC galaxy
within 4$\arcsec$. The faintest NUV sources that are matched to GEEC
galaxies have magnitudes of $\sim$ 24.7, as shown by the black line in
the figure. This figure is shown only as an illustration of the \galex
pipeline processed data. For the remainder of the GEEC analysis we use
Point Spread Function (PSF)-matched magnitudes as detailed in
\textsection \ref{sec-psf}.

\subsubsection{CFHT Wircam K-band observations}\label{sec-kband}

In the 2009A observing season, we obtained observations in the K-band
in the 14 and 21h fields using WIRCAM \citep{wircam} on the Canadian
France Hawaii Telescope (CFHT). Four pointings, each with 33 minutes
of exposure time, were made in each of the two fields. Thus, each
field had a region 30$\arcmin$ X 30$\arcmin$ mapped out. Each pointing
was divided into 80 exposures of 25 seconds each, and was dithered in
a 5 point pattern to fill in the chip gaps. The data were then reduced
and processed by the Elixir
\footnote{http://www.cfht.hawaii.edu/Instruments/Elixir/} and
Terapix\footnote{http://terapix.iap.fr/} pipelines.

\subsubsection{PSF matching and Photometry}\label{sec-psf}

There are at least two approaches to measuring accurate colours for
sources in images with different resolution. One approach is to make a
detailed model of the source in each image with the intent on
measuring its "total" magnitude in each band. This is the approach we
will use with the low redshift data, largely because of the
availability of well understood "model" magnitudes in the
SDSS. However, in the GEEC survey we take the second approach: using
aperture-based photometry after adjusting the images to have the same
PSF.

The original photometry of the CNOC2 survey was not accurate enough to
allow detailed SED-fitting. In \citet{Balogh_cnoccol}, we presented
the results of an extensive campaign to obtain new optical and
near-infrared photometry. In this paper, we follow the same general
scheme for making photometric measurements. While we briefly discuss
this scheme below, we refer the reader to \citet{Balogh_cnoccol} for
the full details.

The original GEEC object detection was done on a combined R(CFH12K)
and $r$(Megacam) 'super' image for each of the CNOC2 patches. This
image was used for detection as it allowed for the greatest depth and
coverage, while closely resembling the original R$_C$ band used in
sample selection. We use this 'super' image only for object detection,
and use each filter (including $r$ and $R$) separately for the
photometry. All the photometric images, including the \galex NUV, FUV
and Wircam K-band images, were aligned to the
corresponding super image using \swarp \citep{Bertin+02} after assuring
that the astrometric alignment was correct. In Table
\ref{table-photometry}, we list each of the instruments and filters
used in the paper. We also list the fraction of galaxies in the GEEC
sample that are covered by each filter. Notice that this simply
measures the fraction of galaxies within the footprint of the data,
rather than the fraction detected.

The \galex images have the largest PSF of any of our images ($\sim$
4.7 $\arcsec$), so we convolve each of our images with a Gaussian
kernel appropriate to obtain a PSF of 4.7 $\arcsec$. We then use
\sextractor v2.5 \citep{sextractor} in two-image mode on each image,
with the unconvolved 'super' $r$ image as the detection image. The
photometry is then measured within 10\arcsec apertures, thus ensuring
that we obtain PSF-matched colors for all the wavebands. Total
magnitudes are derived by computing the CFH12K $R$ or Megacam $r$
\sextractor MAG$\_$AUTO parameter in the unconvolved image and then using
the appropriate aperture color. It is
this photometry we use for the rest of the GEEC analysis. 

Measuring photometry in large apertures could introduce biases due to
crowding from other galaxies. We test how often this occurs by examining
the difference in implied total magnitude in the optical wavebands
when using 3\arcsec apertures in the unconvoled image to that implied
using 10\arcsec apertures in the convolved image. Each of the optical
wavebands provide a consistent picture, that approximately 7$\%$ of
galaxies are affected by crowding. For instance, in the Megacam I
band, we find that 7.3$\%$ of the galaxies have implied total
magnitudes which differ by more than 0.2 mags. Visual inspection of
these galaxies confirm that the errors are due to crowding by nearby
galaxies. However, the fraction of the group galaxies which suffer
from crowding measured in this way (6.9$\%$) is essentially equivalent
to field galaxy fraction (7.4$\%$). The removal of these galaxies from
the sample do not affect the results, and we leave them in the sample
for completeness.

We have assumed that the \galex PSF is a Gaussian profile with a
4.7$\arcsec$ FWHM. However, detailed measurements have shown that the
PSF varies from 4.2$\arcsec$ to 5.3$\arcsec$ (FWHM) in the FUV and NUV
channels, respectively \citep{Morrissey+07}. We examine how these
refinements affect our photometry by examining the Megacam $u$
data. The $u$ data is perhaps the most sensitive to proper PSF
matching to the \galex data as the NUV-$u$ color is a measure of the
dust attenuation at these redshifts. We have convolved the $u$ data
with a 4.2" Gaussian and a 5.3" Gaussian (FWHM). The dispersion in $u$
magnitude implied from comparing these two images is only $\Delta$mag
$\sim$ 0.017, which is smaller than our applied zeropoint
uncertainty. The \galex PSF is also not strickly Gaussian, although
the largest variations are in the wings, which has little effect on
our data. These variations from gaussianity are also a function of
position on the detector. We avoid the extreme effects of
non-gaussianity by restricting to the central 0.6 degrees of the field
of view.

We do not apply Galactic extinction corrections directly to the
photometry. We want to avoid adding extinction corrections to negative
fluxes, which occasionally result from a non-detection. Instead, we
apply a correction to each of the population synthesis models based on
the measured extinction from \citet{dustmaps} for each patch of the
survey. This essentially adds the reddening effect of Galactic
extinction to the models, allowing a direct comparison with the
photometry.

\begin{table}
\begin{center}
\begin{tabular}{| c | c | c | c |}
\hline
Instrument & Filter & Fraction covered \\
\hline
GALEX & $FUV$ & 1.  \\
 & $NUV$ & 1.  \\
CFH12K & $B$ & 0.596  \\
 & $V$ &0.591  \\
 & $R$ &0.594  \\
 & $I$ &0.626  \\
Megacam & $u$ & 0.797  \\
 & $g$ & 0.794  \\
 & $r$ & 0.789  \\
 & $i$ & 0.561  \\
 & $z$ & 0.787  \\
INGRID & $K_s$ & 0.112   \\
SOFI & $K_s$ & 0.210  \\ 
WIRCAM & $K$ & 0.543  \\
\hline

\end{tabular}
\end{center}
\caption[Filters use for SED fitting of GEEC galaxies.]{The filters
  used for SED fitting of GEEC galaxies in the \galex sample. The
  instrument used for the photometry is listed along with the fraction
  of galaxies which are covered by each given waveband. 96.2$\%$ of
  all GEEC galaxies in the \galex sample have either $R$ or $r$ band
  coverage, while 71.2 $\%$ have $K$ or $K_s$ band coverage.}
\label{table-photometry}
\end{table}

We restrict the analysis in this paper to galaxies with $r$ $<$ 22
within the fields with follow-up spectroscopy and to $r$ $<$ 21.5
within the rest of the survey. This has been shown to allow an
unbiased statistical sample with no colour dependence to be generated
\citep{CNOC2,Wilman1}. Due to this restriction, the majority of our
galaxies are well detected in all the photometric bands that have
coverage in the area.

The \sextractor magnitude errors are computed in the standard way,
using the distribution of background noise
\citep{sextractor}. However, when comparing to SEDs it is important
that these errors are not underestimates, and thereby disfavor models
which are good fits. To avoid this, we compute a photometric error
correction by examining the standard deviation of the magnitude
dependent difference between observations of the same galaxy in two
similar filters.  At each half magnitude, we scale the average error
implied by \sextractor to equal that implied using similar filters. We
assume that the scaling factor is due to both filters equally.  This
enables us to estimate, for each filter, a photometric error
correction that depends only on observed magnitude, and this is used
for all galaxies in the sample. This procedure will slightly
overestimate the uncertainty because of real variation in galaxy
types, which causes an intrinsic variation in the color. However, that
variation is expected to be on the order of 0.05 magnitudes
\citep{fukugita}.  We also add zeropoint errors of 0.05 magnitudes to
each of the filters in quadrature, except for NUV, INGRID and SOFI
data, which we have zeropoint errors of 0.03, 0.17 and 0.1 mags,
respectively.

\begin{figure}
\includegraphics[width=8cm]{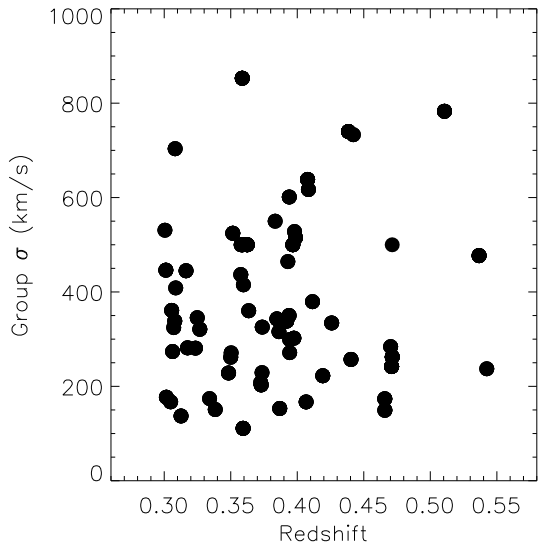}
\caption[The velocity dispersion as a function of redshift for GEEC
  galaxy groups]{The velocity dispersion as a function of redshift for
  GEEC galaxy groups in our three fields with \galex data. We restrict
  our sample to the groups shown, which lie between z=0.3 and
  z=0.55. The typically error in the velocity dispersion is $\pm$ 100 km/s.}
\label{fig-veldisp}
\end{figure}

In order to maintain the unbiased properties of the survey, we
restrict our discussion to galaxy groups that were pre-selected from
the sample of \citet{CNOC2-groups}. However, we do redefine the galaxy
groups using the Wilman \etal (2005) spectroscopy. We follow the
redefinition scheme described extensively in \citet{mcgee} and based
on \citet{Wilman1}. We define galaxy group members as those that are
within 500 \hkpc of the luminosity weighted group centre and have a
line of sight velocity within two times the velocity dispersion of the
group redshift. Using this method, our final sample of 2347 galaxies
between z=0.3 and z=0.55 contains 335 group galaxies. Figure
\ref{fig-veldisp} shows the measured velocity dispersion of our galaxy
groups as a function of redshift. For clarity, we have omitted
velocity dispersion uncertainties, which are typically $\pm$ 100 km/s,
or upper limits, which arise in a few cases when the velocity errors
are larger than the velocity dispersion. Full details of the velocity
dispersion calculations are in \citet{Wilman1}.

In this paper, we call galaxies in our groups to be 'group galaxies'
and we define the 'field' sample to be all other galaxies. The field
sample is not an 'isolated' sample, as it will have many groups
contained within it that are undetected by the sparse CNOC2 redshift
survey. Indeed, we have shown in a previous paper \citep{mcgee}, that
the 'field' sample is very close to a global sample of galaxies and
thereby contains close to the universal fraction of group galaxies
within it.

\subsection{SDSS survey}

The Sloan Digital Sky Survey (SDSS) is a five colour ($ugriz$)
photometric and spectroscopic survey \citep{yorkSDSS}. In this paper,
we make use of the Data Release 6 (DR6), which contains over 790,000
galaxy spectra within approximately 7425 deg$^2$
\citep{SDSSDR6_short}. We restrict our analysis to only galaxies
within the main galaxy sample, which targets almost all galaxies with
$r$ $\leq$ 17.77 \citep{SDSSmainsample}. Some galaxies are left
unobserved because of the finite size of the fiber optic spectra
plugs, which does not allow two galaxies to be observed when they are
within 55$\arcsec$ of each other \citep{Blanton_tiling}.

We use the $ugriz$ $\textsc{modelmag}$, which is designed to give the
most accurate galaxy colour while still being close to a total
magnitude \citep{Stoughton+02}. The SDSS photometry has zeropoint
calibration errors of approximately 0.01 mag in $g$,$r$, and $i$, 0.02
mag in the z band and 0.03 in $u$ \citep{uband}.

 \subsubsection{\galex data}
 
The \galex Medium Imaging survey (MIS) was designed to provide single
orbit ($\sim$ 1500 second exposure) of approximately 1000 square
degrees of the SDSS \citep{Martin+05,Morrissey+07}. We use 1017 tiles
of the MIS that were released to the public via the 3rd General
Release. We retain only objects within the central 0$^\circ$.6 of each
\galex pointing, and resolve overlapping observations by keeping the
one nearest the pointing center. The photometry from the \galex
pipeline is computed using \sextractor in two image mode. In the
pipeline, a detection image is created by dividing a
background-subtracted data image to a corresponding detection
threshold map. \sextractor uses any pixel in the ratio with values
greater than one as possible detections, and measures the
corresponding photometry using a background subtracted image.  
 
We follow the work of \citet{Budavari+09}, who showed that an
effective matching of SDSS galaxies and \galex sources could be made
by matching to the nearest source within 4$\arcsec$. We use all
galaxies within the main galaxy sample of the SDSS as our starting
point. We reduce this sample to those galaxies within 0$^\circ$.6 of a
MIS tile and subsequently match to NUV/FUV sources from the \galex
\sextractor catalogs. We use the MAG\_AUTO photometry, which is close
to a total magnitude. We do not attempt to explicitly match the PSF of
this photometry with the SDSS photometry as we have done for the
UV-optical data of the GEEC sample. Our use of SDSS
$\textsc{modelmag}$ assures that we are measuring total magnitudes in
both regimes, and thus PSF matching is of less importance. The use of
fixed apertures such as was done in the GEEC sample would introduce
significant redshift bias due to the factor of 12 difference in
physical size of an arcsec throughout the low redshift
sample. Further, the resolved galaxies of the low redshift sample make
the PSF becomes less critical.  The final sample of galaxies in the
SDSS/\galex overlap between z=0.01 and z=0.15 is 49663.  87$\%$ of
these galaxies are detected at the 3-$\sigma$ level in $NUV$ and
79$\%$ at the same leve in $FUV$.

\subsubsection{Group finding}

In an attempt to compare the redshift evolution of group properties as
fairly as possible, the primary goal of our low redshift group finder
is to reproduce the selection of our GEEC groups. The method is very
similar to the method used and described in \citet{mcgee} for low
redshift groups.

GEEC groups were selected from a shallower and more sparsely sampled
redshift survey than the SDSS. Thus, we apply two filters to the SDSS
galaxies before applying the group finding algorithm. First, we apply
the same absolute magnitude cut as the original CNOC2 group finder,
M$_R$=-18.5 with an applied evolution correction of 1 mag per unit
redshift. Secondly, we randomly remove half of the remaining galaxies
to replicate the completeness of the CNOC2 redshift survey.

For all of the remaining galaxies, the local density was calculated by
counting the galaxies within a cylinder of 0.33 \Mpch and $\pm$ 6.67
\Mpch line of sight depth centered on it. If the galaxy has fewer than
three neighbours then the cylinder is expanded in every direction by
1.5 and the neighbours recounted. The galaxy is only available to
become a grouped galaxy if it has a higher density than the
background.

The groups are then compiled by beginning at the highest density
galaxy and adding all the galaxies within its cylinder. Next all the
galaxies in each of the cylinders centered on those galaxies are
added. This continues until no more galaxies are added to the
``proto-group''. Members of this group are then used to calculate the
geometric center, redshift and velocity dispersion ($\sigma_v$). Using
these quantities as starting points, galaxies are added or removed
iteratively that within 1.5R$_{200}$, where R$_{200}$ = $\sqrt{3}
\sigma_v/\mathrm{[10 H(}z\mathrm{)]}$, and three times the velocity
dispersion. This process is repeated until convergence.

Using this sample of groups, we now add in all the galaxies that were
randomly removed to reduce the completeness. This allows us to emulate
the GEEC follow-up of the Carlberg et al groups. The geometric center,
redshift and velocity dispersions were again re-computed, now using
all available galaxies. To avoid edge effects, we run the group finder
on the entire SDSS DR6 main galaxy sample, and then subsequently
restrict ourselves to groups within the SDSS/\galex overlap region.

As in the GEEC sample, we call all the galaxies which reside in our
groups using this method 'group galaxies' and describe the rest of the
SDSS sample as 'field galaxies'. But again, due to incomplete
sampling, the 'field' sample has close to the universal value of
galaxy groups. 

This process of finding groups has been shown in \citet{mcgee} to
reproduce a similar halo mass distribution for both GEEC and SDSS
galaxy groups when simulated in mock catalogs. While it is outside the
scope of this paper, in an upcoming work, we will further examine the
group properties at both epochs, including the concentrations and
degree of Gaussianity in velocity distributions (Hou et al., ${\it in~
prep.}$).  The CNOC2 survey (on which GEEC is based) and the SDSS
survey select galaxies only based on $r$ band, with no color
selection.  The similarity in selection and group halo mass function
suggests that the GEEC and SDSS samples can be fairly compared without
a strong bias. However, we should remember that the two redshift
epochs are the result of different surveys and subtle systematic
biases may remain. To minimize the effect of this, we focus
principally on comparing the group and field behavior at a given
epoch, and analyse how these relative differences change between
epochs.

\section{Fitting Spectral Energy Distributions} \label{sec-SED}

UV photometry, while extremely useful for measuring young stars, is
sensitive to attenuation from dust. Early studies analysing UV
photometry from star bursting and rapidly star forming galaxies found
that a reasonable dust correction could be made by assuming that the
galaxy's spectral slope in the UV was proportional to the dust
attenuation \citep{Meurer+95, calzetti}. However, when ultra-violet
light was observed in normal star forming galaxies, and even
relatively quiescent galaxies, it was found that there is no universal
relation between the UV slope and UV dust attenuation
\citep{Bell+02, cortese_dust}. In effect, the UV slope overestimates dust
attenuation for galaxies with significant old populations. This is
particularly important when studying massive galaxies and those in
groups because of their expected old populations.

To overcome these problems, we use all the available photometry
(optical, near-infrared and UV data), to systematically compare
against a sample of models created from the population synthesis code
of \citet{bc03}. This specific method we follow is that of
\citet{Salim+07}, who have earlier used a similar SDSS/\galex overlap
sample.

The general technique involves using a sample of template SEDs with
known galaxy parameters. The templates are fit to the observational
photometry and the galaxy parameters of the best fit model, or the
weighted likelihood of all models, are adopted
as the parameters of the observed galaxy. Below we discuss the
template SEDs that we create using stellar population synthesis
(\textsection \ref{sec-models}) and then discuss the detailed
methodology we use to find the best fitting parameters (\textsection
\ref{sec-fitmethod}).

\subsection{Model Stellar Populations}\label{sec-models}

All our galaxy templates are created using \citet{bc03} models with a
Chabrier initial mass function \citep{chabrier_imf}, assuming a lower
(and upper) IMF mass cutoff of 0.1 (100) M$_\odot$. These templates
use the Padova 1994 evolutionary tracks \citep{padova94_I,
padova94_II, padova94_III, padova94_IV}, while the stellar spectra are
drawn from the BaSeL 3.1 \citep{basel31_I, basel31_II, basel31_IV} and
STELIB \citep{stelib} spectral libraries.

We follow the galaxy parameter ranges used and justified by
\citet{Salim+07}.  The model spectra are produced by randomly
selecting values for a set of variables that control the model
galaxy's age, metallicity, star formation history and dust
obscuration. In particular, the age of the galaxy is randomly selected
from a uniform logarithmic distribution between 0.1 Gyr and the age of
the Universe at the epoch of observation. The metallicity is uniformly
spaced between Z=0.005Z$_\odot$ and 2.5Z$_\odot$, where the canonical solar
metallicity is Z$_\odot$=0.02. 96$\%$ of the model galaxies have
metallicites between Z=0.1-2.5Z$\odot$.

We assume that the star formation histories of these galaxies have a
backbone of exponentially declining star formation rates, and
superimposed onto this backbone history are star formation bursts. In
particular, the backbone rates are randomly chosen with a uniform
distribution in $\gamma$ (SFR $\propto$ $\mathrm{exp}$($-\gamma t$)),
where $\gamma$ is 0 $\leq$ $\gamma$ $\leq$ 1 Gyr$^{-1}$. On this
backbone, we allow bursts that last some time randomly distributed in
duration between 30 and 300 Myr. The strength of the bursts are also
randomly chosen so that during the lifetime of the burst they produce
between 0.03 and 4 times the stellar mass the galaxy had at the onset
of the burst. These bursts are randomly assigned so that there is a
25$\%$ chance a galaxy will undergo at least one burst in a given Gyr.
Each Gyr is independent, and thus many galaxies go longer than 4 Gyrs
without a burst. If the galaxy has formed more recently than a Gyr,
then the chance of it having a burst is prorated. This is the same
burst frequency and strength adopted by \citet{kauffmann_mass} after
examining the burst diagnostic plane of D$_n$(4000)-H$\delta$, and
subsequently adopted by \citet{Salim+07}. Note that, given the length
and frequency of bursts, only $\sim$ 4$\%$ of model galaxies are
undergoing a burst at any one time. This is similar to the 5$\%$ duty
cycle of galaxies exceeding 0.6 dex from the SFR sequence found by
\citet{Noeske+07}.

We adopt the simple two-component dust model of \citet{Charlot+00}. In
this model, young stars are shrouded in the dust associated with their
birth, which has an associated optical depth to the observer of
$\tau_v$. After birth these young stars gradually disrupt and or
dissipate their dust clouds. This is modelled by assuming that only
some fraction, $\mu_v$ of the original optical depth remains 10 Myr
after the stars birth. The $\tau_v$ is drawn from a distribution that
peaks at 1.2 magnitudes of attenuation and runs from 0 to 6 mags. The
$\mu_v$ value runs from 0.1 to 1, peaking at 0.3.

We then create model magnitudes by convolving the \citet{bc03}
resultant spectra with filter curves of \galex NUV and FUV bands as
well as the SDSS $ugriz$ for the low redshift sample. The model
magnitude catalogs are generated at each 0.03 redshift interval from
z=0.02 to z=0.20. When comparing these model catalogs to actual data,
we interpolate between the nearest redshift intervals for each
template to simulate the proper k-corrections. Finally, the distance
modulus is applied to simulate that observed galaxy redshift. In the
GEEC sample, we convolve the spectra with the expected transmission of
\galex NUV and FUV, CFH12K $BVRI$, Megacam $griz$, Wircam $K_s$, and
2MASS $K$. We use the 2MASS filter because both the INGRID and SOFI
$K$ band data were calibrated to give magnitudes on the 2MASS
system. We generate catalogs every 0.05 redshift interval from z=0.3
to z=0.55 and apply interpolated k-corrections and distance moduli.

\subsection{ Fitting Methodology }\label{sec-fitmethod}

The set of model galaxies we have generated by randomly sampling the
allowed parameter space acts as a Bayesian prior to the physical
galaxy parameters. Our goal is to find a resultant probability
distribution function (PDF) for each parameter and for each observed
galaxy. For a given observed galaxy, we find the scale factor, $a_i$,
that gives the minimum $\chi^2_i$ for each model galaxy, $i$, in
Equation \ref{equ-scaling}.

\begin{equation}
\chi^2_i = \sum_X \left(\frac{F_{{\rm obs},X}-a_i F_{{\rm mod}_i,X}}
{\sigma(F_{{\rm obs},X})}\right)^2 
\label{equ-scaling}
\end{equation}

In which, $X$ represents the sum over the 7 bands of low redshift
photometry and the 12 bands of GEEC photometry. $F_{{\rm
obs},X}$ then represents the flux in the Xth band of the observed
galaxy, while, $F_{{\rm mod}_i,X}$ is the Xth band flux of the given
model galaxy. $\sigma(F_{{\rm obs},X})$ is the error of flux in the
$X$th observed band.

We use all available photometric bands for each galaxy. The only
exception arises with GALEX NUV data, and at low redshift, FUV. These
are unique bands for two reasons. First, it is these bands that
largely drive our star formation rates and are thus especially
important. Secondly, these are the only bands for which there are a
significant number of non-detections. Therefore, for galaxies that are
not detected in the original NUV or FUV imaging, we restrict the space
of models to have NUV-r or FUV-r colours redder than the observed
limits. This effectively puts more weight on the NUV and FUV points
given their importance in probing the SFR.  The non-detections , which
make up $\sim$ 20$\%$ of GEEC and $\sim$ 13$\%$ of SDSS galaxies, are
only in the region of 'non-starforming' galaxies, and is largely the
cause of our slight 'overestimate' of the SFR in passive galaxies, as
shown in \textsection \ref{sec-simlgals}.

While the SDSS spectroscopic sample is derived from the $r$ band
photometry, the remaining bands of photometry are sufficiently deep
that accurate colors are obtained for all of the galaxies drawn the
main galaxy sample in our redshift range of interest.  Similarly, with
the GEEC photometry, except for the $z$ band data, all galaxies are
brighter than the expected limits. However, at the faint end, the $z$
band photometric errors become large enough to encompass the models
which are beyond the photometric limit, so we take no explicit action
for these.

The $\chi^2$ value of each model galaxy is then used to define
a weight, exp$(-\chi_i^2/2)$ for that model in fitting the given
observed galaxy. The galaxy parameters of that model are then all
given the weight, w$_i$. Once all models have been assigned a weight,
each galaxy parameter has a PDF constructed by compounding the
associated weights at each parameter value range. We then obtain the
parameter values that correspond to the median of the PDF. Also, we
obtain an approximation to 1$\sigma$ error bars by using 1/4 of the
2.5-97.5 percentile range, and we will quote typical measurement
errors based on this.

The star formation histories we use in the models of Bruzual and
Charlot have a time resolution of 10$^7$ years. Therefore, the final
galaxy parameters at any given epoch are representative only of the
galaxy within the last 10$^7$ years. However, UV star formation rates
are sensitive to any star formation within the last 100 Myrs because
this is the lifetime of the massive stars that produce UV emission. In
other words, a burst of star formation will shine in the UV for 100
Myrs. Accordingly, we average the final 100 Myrs of any simulated
Bruzual Charlot galaxy to produce a parameter that the UV is
tracing. Thus, when we refer to star formation rates produces by this
SED fitting, we are referring to the average star formation rate
within the past 100 Myrs.

\subsection{ Dust enshrouded star formation}

As we show in Appendix \ref{sec-physical}, the SED-fit star formation
rates agree very well with the rates inferred from star forming
galaxies using dust corrected H$\alpha$ emission. However, the
possibility exists that some star forming galaxies could be so
optically thick that they would not emit significantly in H$\alpha$ or
UV. Using star formation rates inferred from UV/optical photometry
like ours and comparing to mid-infrared star formation indicators,
\citet{Salim+09} found that indeed most of the star forming galaxies
were not optically thick.

To obtain a star formation rate from mid-infrared photometry, such as
that taken at 24 $\micron$ by the Spitzer Space Telescope \citep{Werner}, a
correction is made to obtain the total infrared emission emitted by
the galaxy. In the standard manner, this correction is found by
assuming a model, usually empirical, which constrains the SED in the
infrared. There are a variety of models available and \citet{Salim+09}
found that the UV SED fit star formation rates could reproduce the
mid-IR derived star formation rates to within a factor of 2 using the
UV/optical photometry alone when using the models of \citet{Chary+01}
and \citet{Dale+02}. However, when using the more recent templates of
\citet{Rieke+09}, the UV/optical photometry did not accurately
reproduce the dust measured SFR in the sense that Rieke et al method
gave higher star formation rates.  We present analysis of 24 $\micron$
photometry for GEEC galaxies in Tyler et al. (submitted), which uses the
\citet{Rieke+09} templates.  We show there that the amount of dust
enshrouded star formation per galaxy is similar in the groups and in
the field. Since the Rieke et al. templates give higher amounts of
obscured star formation than the \citet{Chary+01} or \citet{Dale+02}
templates, assuming either type of template makes it unlikely that we
are missing a significant population of optically thick star formation
that only exists in one environment. Indeed, \citet{tran_spitzer},
using the templates of \citet{Dale+02}, found that a comparable
fraction of SFR $>$ 3 \Mdot/yr galaxies were found in the group
environment and in the field environment.

\subsection{Simulating galaxy samples}\label{sec-simlgals}

As will be introduced in \S~\ref{sec-ssfr}, the galaxy parameters we
use in this paper are principally the specific star formation rate
(SSFR; the star formation rate divided by the stellar mass) and the
stellar mass. In Appendix \ref{sec-physical}, we have attempted to
quantify the accuracy of these galaxy parameters by comparing them to
parameters obtained using other, largely independent methods. However,
this approach is not fully satisfactory, because the parameters can
only be tested for limited cases, ie. H$\alpha$ measurements provided
only constraints on the low redshift, star forming galaxies. We do not
have a similar test available for the GEEC data. Thus, in this
section, we create a mock sample of observations of galaxies for which
we know the "true" SSFR and stellar mass, which allows us to establish
the legitimacy of our fitting procedure.

\begin{figure*}
\includegraphics[width=\textwidth]{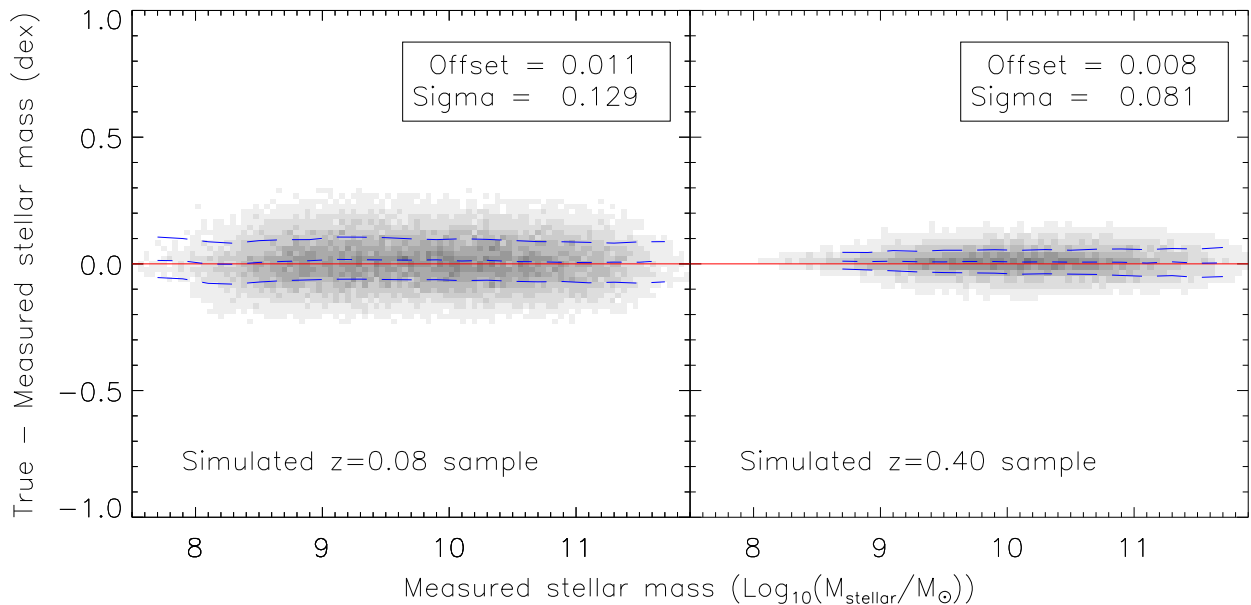}
\caption[Comparison between the true stellar mass and those produced
  by SED fitting simulated galaxies.]{Comparison between the "true"
  stellar masses of a mock set of simulated galaxies and the SED fit
  stellar mass of the mock galaxies. This is shown for both the z=0.08
  simulated sample (left panel) and the z=0.4 sample (right
  panel). The solid red line in both plots shows where the difference
  in measurements is 0. The blue, central dotted line represents the
  running median of the offset, while the two blue, dashed lines show
  the 1 $\sigma$ limits. }
\label{fig-straw-mass}
\end{figure*}

\begin{figure*}
\includegraphics[width=\textwidth]{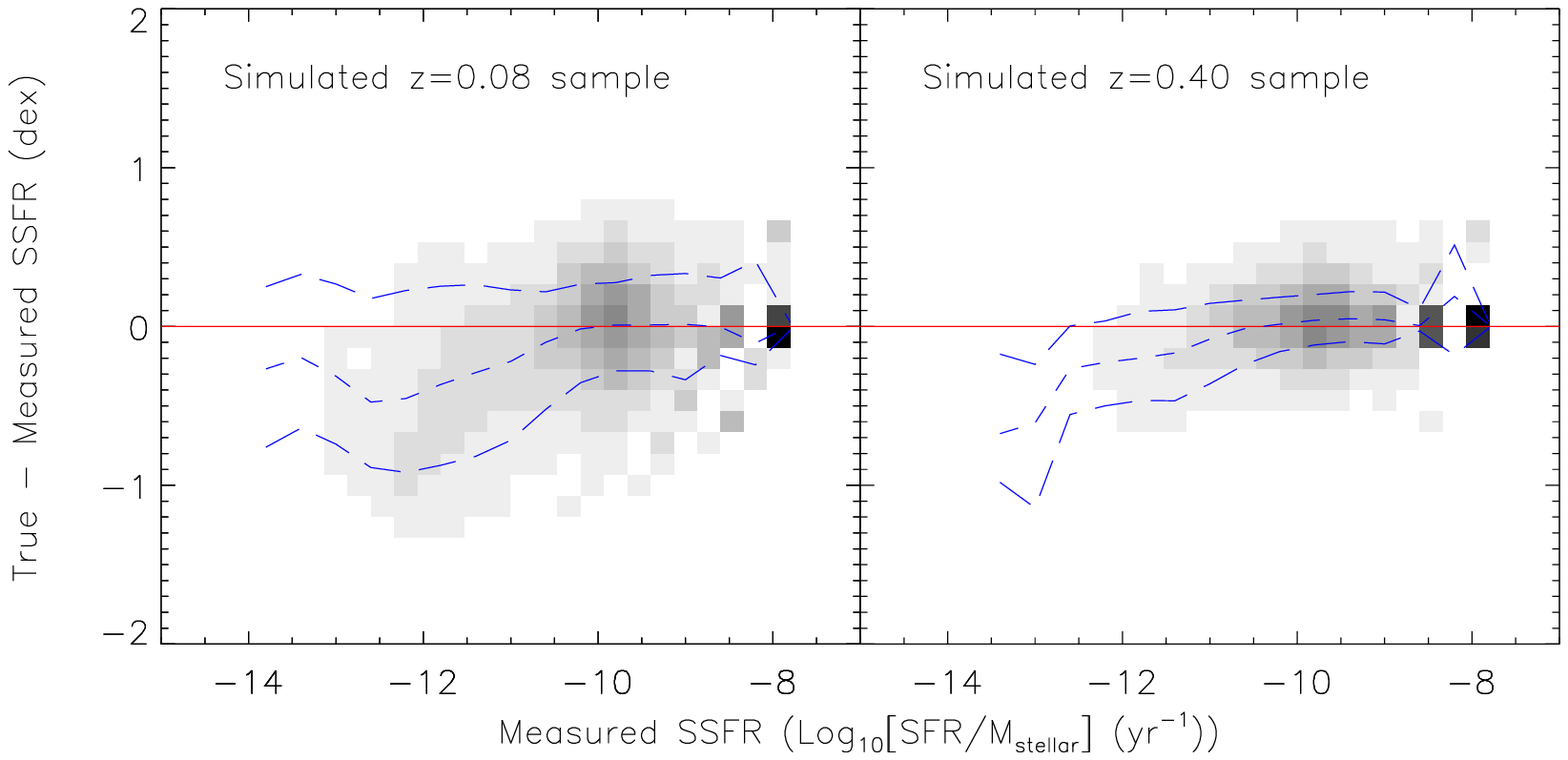}
\caption[Difference between true specific star formation rate and the
  results of SED fitting simulated galaxies.]{The difference between
  the "true" specific star formation rate of our mock galaxies and the
  results of the SED fitting as a function of the SED fit
  results. Again, this is shown for both samples, the z=0.08 (left
  panel) and the z=0.4 (right panel) simulated sample. The blue,
  central dotted line represents the running median of the offset,
  while the two blue, dashed lines show the 1 $\sigma$ limits. }
\label{fig-straw-ssfr}
\end{figure*}

For each survey, we create a sample of 100,000 mock galaxies that are
drawn from the same distribution of dust, metallicity, age and star
formation parameters from which the fitting templates were drawn, as
described in \textsection \ref{sec-models}. However, while the mock
galaxies and fitting templates are drawn from the same distribution,
no single mock galaxy has the exact same parameters as any one of the
fitting templates. The mock galaxies are normalized to have stellar
masses between 10$^{8}$ and 10$^{12}$ \Mdot, and are then placed at
z=0.08 (SDSS mock galaxies) and z=0.4 (GEEC mock galaxies). Mock
observations are then created using the 7 filters of the SDSS sample
and the 12 filters of the GEEC sample. For each mock galaxy, the
photometry in each band is given a random Gaussian error consistent
with the expectations of that bands measurement and zeropoint
errors. In the GEEC sample, we also simulate the heterogeneous nature
of the waveband coverage by removing observations in order to
reproduce a sample with the same fraction of coverage in each
band. However, as in the real method, we always have a minimum of 4
filters of coverage, including \galex NUV and at least one band at $r$
or redder.  Crucially, we also simulate the depth of the \galex NUV
and FUV data. When the mock NUV or FUV observation is beyond the
magnitude limit of our real observations, we restrict to fitting only
models with an NUV-$R$ and FUV-$R$ color greater than the color at the
NUV and FUV observation limit.

In Figure \ref{fig-straw-mass}, we present the stellar mass parameters
that result from SED-fitting the mock set of simulated galaxies in
both surveys. We show the difference between the "true" stellar mass
and those obtained by SED fitting the mock observations as a function
of the SED masses. The fitting recovers very well the true stellar
mass, with only a nominal systematic offset and no mass dependent
deviations in the adequacy of the recovery. Note that the z=0.4 sample
has a smaller scatter ($\sigma$ = 0.081 dex) than the z=0.08 sample
($\sigma$ = 0.129 dex).  This is likely because we restrict the age of
galaxies in the models to be less than the age of the Universe at that
epoch, which means that the z=0.4 sample has less cosmic time for wide
variations in star formation histories.

In Figure \ref{fig-straw-ssfr}, we show the SSFR results produced from
SED fitting the mock galaxies. Unlike the stellar mass comparison, we
see a systematic deviation from the true SSFR that depends on the
SSFR. At the star-forming end, greater than $\log_{10}$(SSFR)$ =
-10.75$ at z=0.08 and $\log_{10}$(SSFR)$ = -11$ at z=0.4, the SSFR is
reproduced correctly with small scatter ($\sim$ 0.25 dex at low
redshift and $\sim$ 0.2 dex at $z=0.4$).  However, below these points,
the SED fit SSFR seems to systematically overestimate the true SSFR by
$\leq$ 0.3 dex. This behavior is a direct result of the depth of the
\galex NUV data. As we have mentioned, when the NUV magnitude is
beyond our completeness, we simply restrict model space to those
models that have $(NUV-R)$ colours redder than that. Thus, the SSFR
is essentially an average of the remaining models, which leads to a
slight overestimate of the SSFR. This is indicative that our priors
for low star formation rate galaxies are not correct. However, in this
paper we restrict ourselves to analysis only of the high star
formation rate galaxies and the fraction of these; thus, our slight
overestimate of the SSFR has no impact for the present analysis and we
do not try to develop more rigorous priors. Also, notice in Figure
\ref{fig-straw-ssfr} that there is a population of galaxies undergoing
an intense burst of star formation and thus have very high SSFRs
($\sim$ 10$^{-8}$ yr$^{-1}$).

Our mock sample fitting has led to two important points. First, as
also evidenced by our comparisons with published stellar masses in
\textsection \ref{sec-physical}, we are able to accurately recover
stellar masses to the level of $\sigma$ = 0.2 dex over the full range
of mass within the constraints of the models. Second, we have shown
that we can recover SSFR for star forming galaxies, but that we may be
overestimating the SSFRs of non-star forming galaxies. This is an
important point, since the \citet[B04]{brinchmann_sfr} star formation
rates did not allow us to estimate our accuracy at low SFR. With these
findings in mind, in this paper, we avoid making a detailed analysis
of the rates of low star forming galaxies and instead set a threshold
for the division of active and passive galaxies at $\log_{10}$(SSFR)$
= -11$, essentially lumping all passive galaxies together.

\section{Results } \label{sec-results}

\subsection{SSFR-M$_\odot$ plane}\label{sec-ssfr}

The star formation rate of a galaxy is a key indicator of galaxy
evolution, and for active galaxies there is a strong correlation
between star formation and stellar mass. Many authors,
\citep[eg.][]{kennicutt+83, scalo, brinchmann_sfr}, have thus examined
the specific star formation rate (SSFR) which is defined as the star
formation rate normalized by the total stellar mass of the galaxy:

\begin{equation}
\mathrm{SSFR} \equiv \frac{\mathrm{SFR}}{\mathrm{M}_{\mathrm{stellar}}}
\end{equation}

\begin{figure*}
\includegraphics[width=\textwidth]{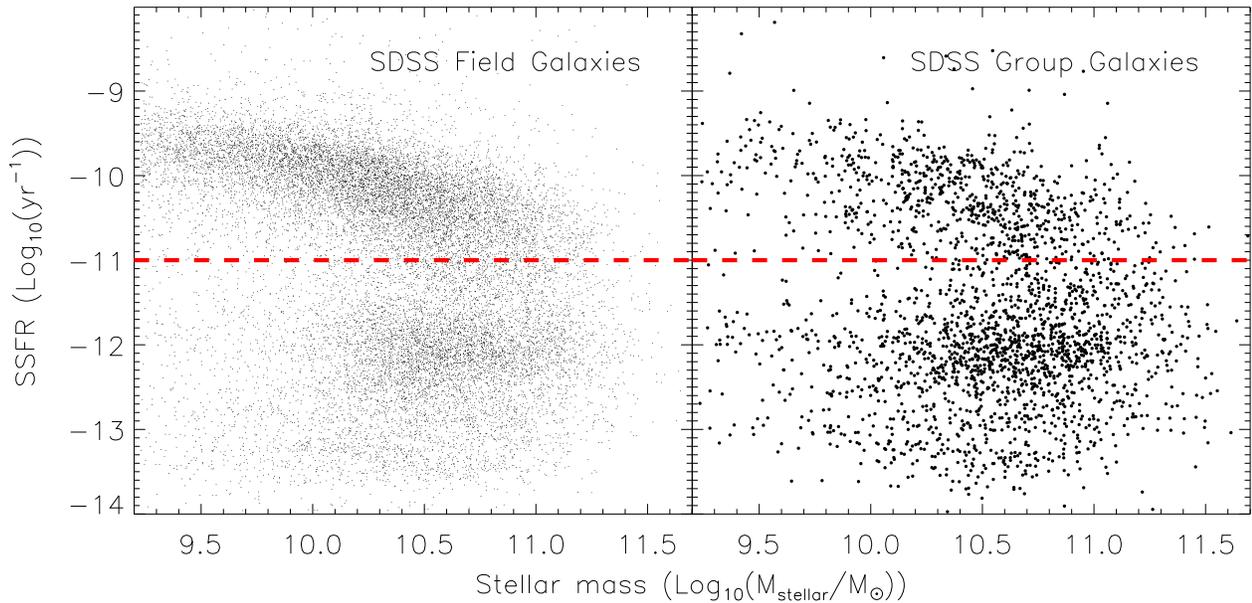}
\caption[Specific star formation rates of both the group and field
  galaxies in the SDSS survey]{Specific star formation rates of both
  the group (right) and field (left) galaxies in the SDSS survey
  (median redshift 0.08) as a function of stellar mass. The dashed,
  red line corresponds to $\log_{10}$SSFR = 10$^{-11}$yr$^{-1}$, which
  is our division between active and passive galaxies. The typical
  uncertainty on SSFR is 0.31 dex, while the stellar mass
  uncertainty is 0.28 dex.}
\label{fig-sdss_sfr}
\end{figure*}

This is a useful approach because it quantifies the current SFR with
respect to the past SFR \citep{kennicutt_94}. For instance, a value of
SSFR of 10$^{-9}$ yr$^{-1}$ means that if the galaxy maintains its
current SFR for 10$^{9}$ yr, then it will double its stellar
mass. SSFR can then easily be converted into a ``birthrate'', $b$,
defined as $b$ = SSFR $\times$ $t$, where $t$ is the age of the
universe. A value of $b$ greater than 1 means the galaxy is forming
stars faster than it has in the past. Of course, this is an
approximation, as the galaxy may have been forming stars for only a
fraction of the age of the universe, or had a significant amount of
the formed stellar mass returned to the ISM.

It is with this intuition that we present Figure \ref{fig-sdss_sfr},
which shows the SSFR as a function of the galaxy stellar mass for
galaxies in the SDSS sample. We have two panels: one showing the field
galaxies, and one showing the group members. It should be noted that
the points each represent individual galaxies, but that the galaxies
have not been weighted by 1/V$_{\mathrm{max}}$. Thus, especially
galaxies in the bottom left corner (those of low mass and SSFR), are
systematically under-represented. However, when defining passive
fractions or other quantities we always account for these weights.

\begin{figure*}
\includegraphics[width=\textwidth]{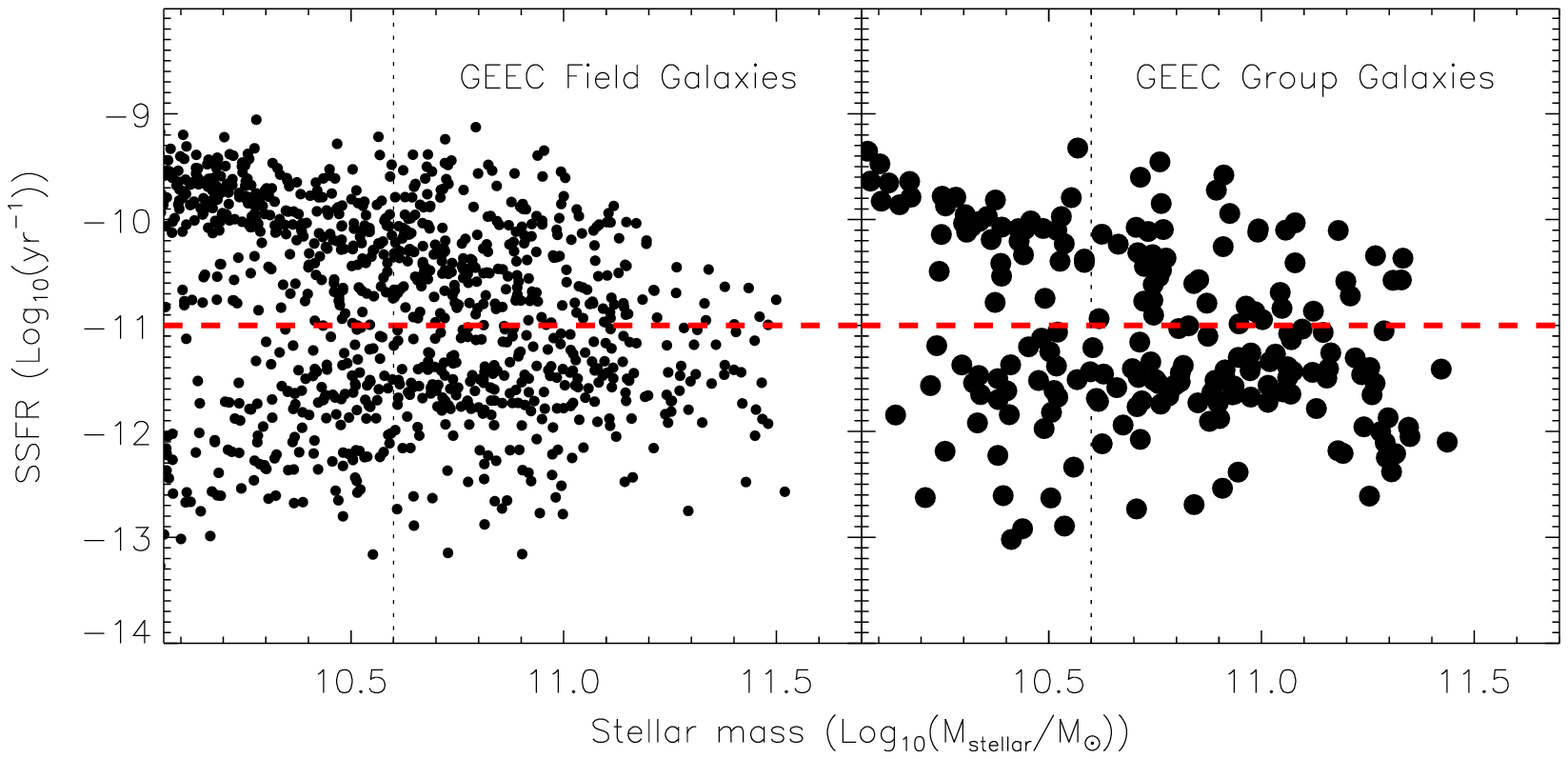}
\caption[Specific star formation rates as a function of stellar mass
  for both group and field galaxies in the GEEC survey.]{Specific star
  formation rates as a function of stellar mass for both the group
  (right) and field (left) galaxies in the GEEC survey (redshift =
  0.4). The dashed, red line corresponds to $\log_{10}$SSFR =
  10$^{-11}$yr$^{-1}$, which is our division between active and
  passive galaxies. The vertical, black, dotted line corresponds to
  the stellar mass limit (M$_{\mathrm{stellar}}$ = 3.9 $\times$
  10$^{10}$) at the highest redshift of the sample (z=0.55). Using
  1/V$_{max}$ weighting, we can obtain a complete sample to
  M$_{\mathrm{stellar}}$ = 1.44 $\times$ 10$^{10}$, but the scatter
  plot does not reflect these weightings. The typical uncertainty on
  SSFR is 0.28 dex, while the stellar mass uncertainty is 0.19 dex.}
\label{fig-geec_sfr}
\end{figure*}

Examining this figure, it is apparent that galaxies appear bimodal in
the SSFR-\Mstellar plane. They tend to cluster as part of either a
group of passive galaxies at SSFR $\sim$ 10$^{-12}$, or a group of
star forming galaxies at SSFR $\sim$ 10$^{-10}$. This behavior is
obvious in both the group and field galaxies. The galaxies clustered
at $\sim$ 10$^{-10}$ has been called by other authors the ``main
sequence of star-forming galaxies'' \citep{Noeske+07}. These galaxies
have a birthrate, $b$, of approximately 1, which could suggest that
they have been forming stars at about this rate for their lifetime.

It is important to remember that, as shown in \textsection
\ref{sec-simlgals}, the low SSFR rates were actually slight
overestimates, meaning that the cluster of points at SSFR $\sim$
10$^{-12}$ is unlikely to be caused by systematically low SSFR
measurements. In other words, this 'main sequence' is not due to an
insensitivity to star formation rates just underneath the
sequence. This is a crucial point as the work of \citet{Noeske+07} was
based principally on emission lines, which are relatively insensitive
to low star formation rates. We stress that because our star formation
rates are based on SED-fitting, they are complete as a function of
stellar mass, with no selection on the star formation rate. Thus,
similar to the behaviour seen in galaxy colours, the bimodality of
galaxy populations seems to be a fundamental property.

Many authors have attempted to quantify this ``main sequence of star
forming galaxies'' through linear or model fitting \citep[][Gilbank et
al., submitted]{Noeske+07, Peng+10}.  Because our principal goal in
this paper is to examine the differential evolution of group and field
galaxies, we will avoid these kinds of parametric fits. However,
visual inspection of this figure does yield at least two interesting
features. First, the `main sequence' in Figure \ref{fig-sdss_sfr}
appears to have a steep slope, such that low mass galaxies have
significantly higher SSFRs than do more massive galaxies. If we assume
that this star forming sequence of galaxies has always been star
forming, then the massive galaxies have either formed earlier than
less massive galaxies or have more rapidly declining star formation
histories relative to less massive galaxies.  Secondly, the slope and
position of the group galaxies does not appear wildly different from
the field galaxies. We will revisit this issue later.

We now move to higher redshift to examine the GEEC sample at z $\sim$
0.4. In Figure \ref{fig-geec_sfr}, we show the
SSFR-M$_{\mathrm{stellar}}$ plot for GEEC galaxies, separated into
group and field environments. We only plot the stellar mass region in
which we can create a statistically complete sample by applying the
spatial, magnitude and 1/V$_{\mathrm{max}}$ weights. As in the
previous figure, we see that it appears that galaxies form a sequence
of star forming galaxies. Again, the sequence appears to have a tilt,
such that massive galaxies have lower SSFRs than lower mass galaxies.

\begin{figure}
\includegraphics[width=8cm]{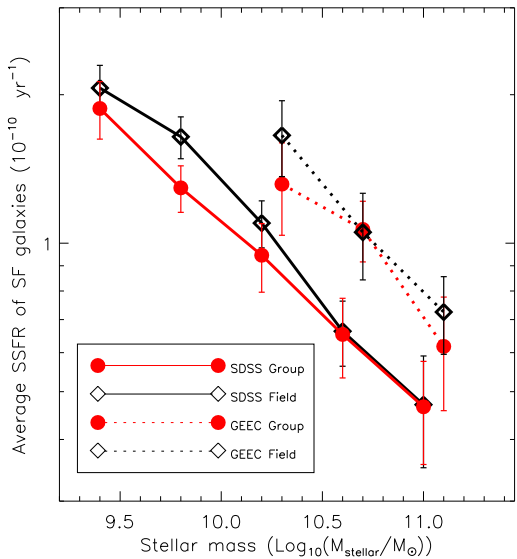}
\caption[Average SSFR of star forming galaxies in the group and field
of SDSS and GEEC surveys]{Average SSFR of star forming galaxies in the
group and field of SDSS and GEEC surveys. Star forming galaxies are
defined as being above $\log_{10}$SSFR $>$ 10$^{-11}$yr$^{-1}$, the red
dashed line in Figures \ref{fig-sdss_sfr} and \ref{fig-geec_sfr}. The
error bars represent the error in the mean.}
\label{fig-both_ssfr}
\end{figure}

As was the case for the SDSS galaxies, it does not appear, via visual
inspection of Figure \ref{fig-geec_sfr}, that the specific star
formation rates of star forming galaxies are very different in groups
compared with the field, in the GEEC sample.  To quantify this and
to evaluate the evolution with redshift, we present Figure
\ref{fig-both_ssfr} which shows the average SSFR of galaxies
classified as star forming in both surveys. To avoid parametric forms,
we simply assume that star forming galaxies are those with
$\log_{10}$SSFR $>$ 10$^{-11}$yr$^{-1}$. As shown by the red, dotted
line in Figures \ref{fig-sdss_sfr} and \ref{fig-geec_sfr}, this
appears to divide the star forming and passive galaxies at both
redshifts. Further, a galaxy with this SSFR is forming stars at a
fraction of its past average. This division is also useful because it
implies that a galaxy forming stars at this SSFR with a mass of our
GEEC stellar mass limit ($\sim$ 10$^{10}$ \Mdoth) would form stars at
a rate of 0.1 \Mdoth/yr. Thus, the depth of UV photometry easily
discriminates between star forming and passive galaxies.

Focusing first on the general field trends, shown by the black lines
and symbols, we see that both the field samples in the GEEC survey and
in the SDSS survey show steep trends with stellar mass. This behaviour
was pointed out in the scatter plots earlier, but this confirms that
the behavior remains when accounting for volume weighting of the
surveys.  We stress that this behavior is also not a result of a cut
in star formation rate, as we are complete to the stellar mass limits
shown. We notice that there is a significant offset in the
SSFR-\Mstellar relation between surveys for these star forming
galaxies, such that star forming galaxies at redshift z=0.4 were
forming stars at higher rates ($\sim$ 0.25 dex in SSFR) than those at
z=0.08. This offset is mass-independent for the mass ranges that are
probed by both surveys ($>$ 10$^{10}$\Mstellar).

Now, focusing on the group and field values within a given redshift,
we see that, within the statistical uncertainties, there is no
apparent difference. Thus, at least for the star forming galaxies,
from z=0.4 to z=0.08, we see a general lowering of the level of star
formation in galaxies with no apparent mass dependence regardless of
their environment.

\subsection{Fraction of passive galaxies}

\begin{figure*}
\includegraphics[width=\textwidth]{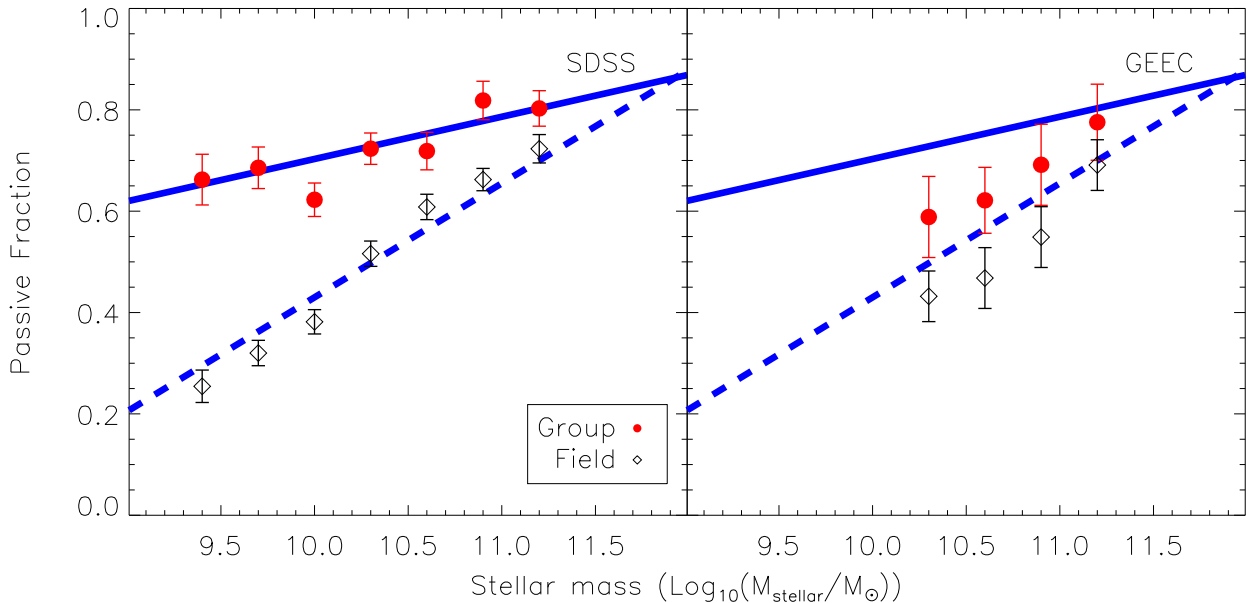}
\caption[The fraction of passive galaxies in the group and field of
  SDSS and GEEC surveys.]{The fraction of passive galaxies in the
  group and field of SDSS (left panel) and GEEC (right panel)
  surveys. Passive galaxies have SSFRs less than
  10$^{-11}$yr$^{-1}$. The solid blue line is an approximation to the
  mass dependence of the SDSS group galaxies, while the dashed blue
  line is an approximation to the SDSS field galaxies. For
  illustrative purposes, the same lines have also been reproduced in
  the GEEC panel. The error bars represent the error in the mean.}
\label{fig-both_fractions}
\end{figure*}

We have concentrated principally on the sequence of star forming
galaxies and its position on the SSFR-\Mdotspace plot. We now shift to
look at what fraction of galaxies are within this sequence. As above,
we define passive galaxies as those that have $\log_{10}$SSFR $<$
10$^{-11}$yr$^{-1}$.  Figure \ref{fig-both_fractions} shows the
passive fractions as a function of stellar mass for the group and
field galaxies in both the SDSS survey and the GEEC survey. At both
redshifts, we see that at all stellar masses probed, groups have a
higher fraction of passive galaxies than the field. Given that this is
comparing group galaxies with field galaxies at fixed stellar mass and
redshift, it is strong evidence that group galaxies have star
formation prematurely truncated by the group environment.

To guide the eye in Figure \ref{fig-both_fractions}, we show a simple
least square fit to each of the group and field passive fraction in
SDSS. These two lines have very different slopes, implying that there
is a mass dependent difference in the truncation mechanisms in the
groups and the field. This is most interesting when comparing to the
GEEC sample. The SDSS lines have been re-drawn on the GEEC panel as
well. This shows that the group and field galaxies at z=0.4 have a
mass dependent fraction with a similar slope to the SDSS field
galaxies. Thus, the evolution in groups must be mass-dependent to end
with the correct fraction of passive galaxies in groups at z=0. This
may be expected because in groups, the low mass galaxies are more
likely to be satellites rather than central galaxies, and therefore,
might experience the loss of their gas reservoir. These results are
intriguing given our results in Figure \ref{fig-both_ssfr}, which
showed that the average SSFR of star forming galaxies was similar in
group and field at a given redshift. Galaxies must move quickly from
being classed by our definition as 'star-forming' to 'passive' to
cause the growing fraction of passive galaxies observed, while keeping
the average SSFR of star forming galaxies unchanged.

\section{Discussion} \label{sec-discussion}

\subsection{Accretion model}

\begin{figure}
\includegraphics[width=8cm]{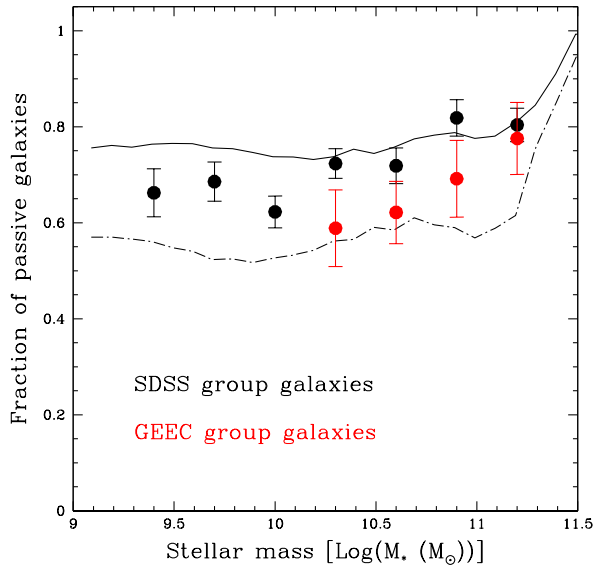}
\caption[Comparison of accretion model with passive fraction in
  groups]{Comparison between the passive fraction of GEEC and SDSS
  group galaxies and a simple accretion model for environmental
  effects. Passive galaxies were determined via their specific star
  formation rates. The accretion model assumes that a galaxy become
  passive 3 Gyrs after it falls into a halo of at least 10$^{13}$
  \Mdot. The solid black line is the prediction at the median redshift
  of SDSS, z=0.08, while the dashed black line is the model at z=0.4.}
\label{fig-accretionmod}
\end{figure}

In \citet{mcgee_accretion}, we introduced a simple approach for
relating environmental effects to a galaxies accretion history. In
this model, galaxies become 'environmentally affected' some time,
\Ttrunc, after they fall into a host halo with a mass greater than
\Mtrunc. It is important to note that our use of 'environmentally
affected' does not necessarily mean, for example, red galaxies,
because galaxies can also become red through internal processes. We
are attempting to reproduce the {\it differential} effect of the group
environment with time rather than making predictions for the group or
field alone. Our model is based on the semi-analytic work of
\citet{font}, but uses only the accretion history and stellar mass
outputs. Our approach does not depend on the detailed star formation
histories predicted by the \citet{font} method. In Figure
\ref{fig-accretionmod}, we show the results for a model where \Ttrunc=
3 Gyrs and \Mtrunc=10$^{13}$ \Mdot. This shows the predictions for the
environmentally affected fraction at z=0.08 and z=0.4, as well as the
observed group passive fractions at these redshifts. To match the
overall SFR-determined passive fraction this model has a higher mass
threshold than was shown previously by \citet{mcgee_accretion}, which
had been determined to match optical 'red' fractions. However, as
shown in \citet{mcgee_accretion}, the rate of evolution in these
models is principally determined by \Ttrunc.  While this model is
simple, it is interesting to note that it predicts approximately the
correct evolution in the passive fraction. Notice that at
M$_{stellar}$ = 10$^{10.3}$ \Mdot, the evolution between the two
redshift epochs in passive fractions is observed to be $\sim$
0.15. The model with a \Ttrunc of 3 Gyr agrees with this observed
evolution.  In contrast, a model with the same \Mtrunc but with
\Ttrunc= 1 Gyr predicts only an increase of 0.03. A timescale this
short is disfavoured by the data, assuming that all galaxies in groups
truncate with this timescale. As we have argued in
\citet{mcgee_accretion}, the most powerful argument against a higher
\Mtrunc\ is simply the observed difference in group and field passive
fractions. A mechanism that occurs in the halo mass and timescale
regime of our successful model could be strangulation, the relatively
gentle process of removing the outer hot halo of infalling galaxies.

This accretion model is simplified in many ways and should only
serve as a guide that the overall level of evolution seen in the data
implies a long timescale. The galaxy formation model does not include
the production of intragroup mass, which can be a significant fraction
of the total stellar mass \citep{mcgee_supernova}. Further,
\citet{kim_clustering} have shown that the inclusion of
satellite-satellite merging and satellite-disruption to create
intragroup mass improves the model's agreement with the luminosity
dependent correlation function. The inclusion of these processes would
likely create mass-dependent evolution since these process
preferentially occur in low mass galaxies.

\subsection{The role of environmental processes }

We have seen that the sequence of star forming galaxies evolves with
redshift, but that it is independent of environment. However, given
that the fraction of star forming galaxies has such a strong
environment dependence, we can ask the question: Are our results
consistent with all low mass passive galaxies residing in massive
halos? It is important to remember that our field sample is
essentially the 'total' sample rather than an 'isolated' sample. The
models of \citet{font} show that 42$\%$ of galaxies are in halos above
10$^{12.75}$\Mdotspace at z=0 while at z=0.4, there are 26$\%$ of
galaxies in these halos. Assuming that all of the galaxy groups have
properties given by our sample, and the total population has the value
of our 'field' data, we can infer the properties of 'isolated'
galaxies, those in halos below 10$^{12.75}$\Mdot. We can relate the
group and field passive fractions ($f_{\mathrm{passive,group}}$,
$f_{\mathrm{passive,field}}$) to the isolated passive fraction
($f_{\mathrm{passive,isolated}}$) using the fraction of galaxies in
groups and those which are isolated fraction at that redshift
($F_\mathrm{G}$, $F_\mathrm{I}$), as

\begin{equation}
f_{\mathrm{passive,isolated}}=
\frac{f_{\mathrm{passive,field}}-F_\mathrm{G}*f_{\mathrm{passive,group}}}{F_\mathrm{I}}.
\end{equation}

In this way, we see that the implied isolated passive fraction at
M$_{stellar}$ = 10$^{11}$ \Mdotspace is 0.50
($\frac{0.55-0.26*0.69}{0.74}$) at z=0.4 and is 0.56
($\frac{0.66-0.42*0.80}{0.58}$) at z=0.  Thus, there is an intrinsic
high passive fraction at this mass in isolated galaxies, but most of
the evolution in the field between z=0.4 and z=0 is driven by the
evolving group passive fraction. At M$_{stellar}$ = 10$^{9.4}$ \Mdot,
for which we only have data at z=0, the implied true isolated fraction
is essentially zero ($\frac{0.27-0.42*0.66}{0.58}$).  At lower mass,
the field passive fraction appears to be entirely due to galaxy
groups.

From this emerges a general picture in which massive galaxies have a
high fraction of passive galaxies regardless of their environment,
while low mass galaxies are essentially only passive in groups. This
implies that massive galaxies must have an internal mechanism, or at
least one that does not depend on environment, which plays a large
role in the galaxy properties. On the other hand, low mass galaxies
are essentially all star forming unless they are in a group or cluster
environment. There have been models in which this internal mechanism
is proposed to be the resultant heating caused by the fueling of
active galactic nuclei \citep{springel_agn, hopkins_agn}.

It is perhaps surprising that we find most, if not all, of the low mass,
passive galaxies can be accounted for if we assume they are
only created in groups.  This argument is based on CDM predictions of
halo abundances and, while reasonably robust, does not rule out the
possibility that some isolated galaxies in the real Universe may be
passively evolving.  Indeed, there is some evidence for such galaxies
\citep[eg.,][]{MalinCarter, colbert, zhu}
although these tend to be more massive than our lowest mass
galaxies. However, there is observational support in the literature for
this claim. In \citet{baldry06}, only 5$\%$ of galaxies at the lowest
stellar mass (10$^{9}$ \Mdot) and in the lowest density region are on
the red sequence. These authors also showed that, while on average the
lowest density region hosts more isolated galaxies, there are a
significant number of galaxies which are in group scale halos but
which nonetheless appear to be in low density environments via their
density indicator. Further, \citet{haines_physical}, using the fourth
data release of the Sloan Digital Sky Survey, find that none of the
$\sim$ 600 galaxies in their lowest luminosity bin (-18 $<$ M$_r$ $<$
-16) and lowest density quartile are passive. Unfortunately, the
literature can be misleading, because it is often convenient to call
the lowest density bin of an observational sample 'isolated'. While
this label is fine for the bulk properties, there can often be
interlopers from groups or clusters, as \citet{baldry06} have
shown. It has also been shown that a significant population of
galaxies which appear isolated may be members of a 'backsplash'
population, which have been through a massive halo and emerged on the
other side \citep{gill_backsplash}. However, even after accounting for
interlopers and backsplash galaxies, in apparent contradiction with
\citet{haines_physical}, \citet{wang_reddwarf} have claimed that
30$\%$ of low luminosity (M$_r$ $>$ -17) SDSS are red and
isolated. The absence of a tight red sequence in these galaxies makes
the division between red and blue galaxies difficult and is determined
by an extrapolation of the luminosity dependent division of luminous
galaxies. Given the lack of a tight red sequence, it is unclear
whether a red $g$ - $r$ color as defined by these cuts is necessarily
a passive galaxy and not a dust-reddened one. We note that the results
of Haines et al., who determine passivity from the absence of
H$\alpha$ emission, are less sensitive to these problems.

It is worth noting that galaxy formation models, such as the
semi-analytic models of \citet{Bowermodel} and \citet{croton}, also
predict that low mass passive galaxies are exclusively satellites of
more massive galaxies in groups and clusters. In these models, the
cessation of star formation in a galaxy occurs primarily through AGN
feedback or satellite related processes.

One important question which arises is whether our results can be
explained if galaxy evolution in groups is simply given a
'headstart'. That is, the galaxies within groups follow the same
evolutionary path as isolated galaxies, but because of their dense
environments they simply formed earlier and are thus further evolved.
Simulations of dark matter haloes do show that there is a
relationship at fixed halo mass between the age of the halo, or its
epoch of formation, and its environment, such that haloes in a more
biased environment form earlier \citep{gaospringelwhite, wechsler06,
Maulbetsch,limogao}. Unfortunately, determining whether this bias
dependent formation time of haloes at fixed $\it{halo\ mass}$ leads to
an environment-dependent formation time of galaxies at fixed
$\it{stellar\ mass}$ requires a more detailed understanding of galaxy
formation. However, as we have seen in Figure \ref{fig-both_ssfr}, the
average star formation rate at a given redshift is similar in the
group and the field, but is significantly different at different
redshifts. If galaxies simply ran out of gas, group galaxies would
have different average star formation rates from field galaxies at the
same redshift if they had a 'headstart'.

Further, it appears that most 'local' effects like AGN correlate
strongly with stellar mass \citep{vonderlinden}. But controlling for
stellar mass, we still see a higher passive fraction in groups. Although
more work still needs to be done, this is strong evidence that
galaxies have their star formation truncated in groups by an
environment specific process.

\section{ Conclusion}\label{sec-conclusions}

We have fit spectral energy distributions to galaxies in two surveys,
SDSS and GEEC. These SEDs use high quality, space based ultraviolet
imaging along with optical, and near infrared for GEEC, photometry. We
have compared this photometry to large suites of stellar population
synthesis models to determine star formation rates and stellar
masses. This method nicely reproduced alternative methods of measuring
both star formation rates and stellar masses. By examining the
results, we conclude the following.

\begin{itemize}
\item Star forming galaxies of all environments undergo a systematic
  lowering of their star formation rate between z=0.4 and z=0.08
  regardless of mass. 

\item The star formation properties of star forming galaxies, as
  measured by their average specific star formation rates, are the
  same in the group and field environment at fixed redshift. 

\item The fraction of passive galaxies is higher in groups than the
  field at both redshifts. However, the difference between the group
  and field grows with time and is mass dependent, in the sense the
  the difference is larger at low masses. 

\item Low mass galaxies at z=0 have group and field passive fractions
  that can be explained if passive galaxies only exist in groups.

\item The evolution of passive fractions in groups between z=0.4 and
  z=0 is consistent with an accretion model in which galaxies are
  environmentally affected 3 Gyrs after falling into a 10$^{13}$ \Mdotspace
  halo/group.

\end{itemize} 

These results present a consistent picture of environmental effects
when taken along with our measurements of quantitative morphology in
\citet{mcgee}. In that paper, we showed that the fraction of disk
galaxies is higher in the field at both redshift and the difference
grows larger with time. Also, we found that there was no indication
that the disk scaling relations were different in the field or groups.

These results all suggest that only a fraction of galaxies in groups
must be actively truncated at any given time. If the mechanism is a
quick one, like ram pressure stripping, then the truncation time might
be short enough that only a small fraction of group galaxies are
affected at a given time. This would allow the bulk of the star
forming galaxies to remain unchanged but still allow the observed
evolution. However, ram pressure stripping is likely not effective in
galaxy groups.

Based on the timescale suggested by our accretion model, strangulation
seems like a suitable candidate for environmental mechanisms. However,
it is unclear if strangulation can allow galaxies to remain apparently
unaffected for some time, thereby appearing to act only on a fraction
of group galaxies at a time. In semi-analytic models, strangulation
produces too many 'green' galaxies, which would likely alter the disk
and star formation properties \citep{font, Balogh_cnoccol}.

These results suggest that further constraints can be applied in two
ways. First, the detailed study of individual galaxy groups and the
orbits of their constituent galaxies within them can determine if ram
pressure stripping of the cold gas is a viable mechanism. Secondly,
the continued hunt for elusive `green' transition galaxies, perhaps at
high redshift where the accretion rates of galaxies into groups is
higher \citep{mcgee_accretion, geec_hiz}, will
determine the viability of gentle, strangulation like mechanisms.

\section*{Acknowledgments}

We thank the referee for a thorough and constructive report. MLB and
LCP acknowledge support from NSERC Discovery Grants. We would also
like to thank the original CNOC2 redshift survey team, who allowed us
access to their unpublished redshifts. This paper is based on
observations made with the NASA Galaxy Evolution Explorer under GO
program 37 and archival data from the 3rd General Release. GALEX is
operated for NASA by California Institute of Technology under NASA
contract NAS-98034. This work is based in part on data products
produced at the TERAPIX data center located at the Institut
d'Astrophysique de Paris. Some of the data presented in this paper
were obtained from the Multimission Archive at the Space Telescope
Science Institute (MAST). Support for MAST for non-HST data is
provided by the NASA Office of Space Science via grant NAG5-7584 and
by other grants and contract. Observations used in this paper were
obtained with WIRCam, a joint project of CFHT, Taiwan, Korea, Canada,
France, at the Canada-France-Hawaii Telescope (CFHT) which is operated
by the National Research Council (NRC) of Canada, the Institute
National des Sciences de l'Univers of the Centre National de la
Recherche Scientifique of France, and the University of
Hawaii. Funding for the creation and distribution of the SDSS Archive
has been provided by the Alfred P. Sloan Foundation, the Participating
Institutions, the National Aeronautics and Space Administration, the
National Science Foundation, the US Department of Energy, the Japanese
Monbukagakusho, and the Max Planck Society.

\bibliography{ms}

\appendix

\section{Understanding the physical parameters}\label{sec-physical}

In this appendix, we attempt to determine how well our SED fit galaxy
parameters reproduce the values given by alternative methods. We will
compare our SDSS stellar masses and star formation rates with the
publicly available results of \citet[K03; for stellar
masses]{kauffmann_mass} and \citet[B04; for star formation
rates]{brinchmann_sfr}. We also compare our GEEC stellar masses to
those determined primarily from $K$ band data by \citet{Balogh_smass}.

\subsection{SDSS comparisons: Stellar Masses}

The method of fitting stellar masses used by K03, is the philosophical
forerunner of the Bayesian method we use here, with much of the
formalism explained in that paper. The authors use a large suite of
galaxy models created using stellar population synthesis, and find
that they can constrain the $z$ band mass to light ratios using the
4000-\AA\ break (D$_n$(4000)) and the Balmer absorption line index
H$\delta_A$. By then scaling by the observed $z$ band photometry, they
determine the galaxy's stellar mass. Thus, despite the similar
methodology, the K03 stellar masses are largely determined by the
spectral features, which determine the star formation history and thus
the mass to light ratio. In contrast, our stellar masses are
determined by using the UV light to constrain the star formation
histories and then the optical/NIR photometry to scale to a stellar
mass.

\begin{figure}
\includegraphics[width=8cm]{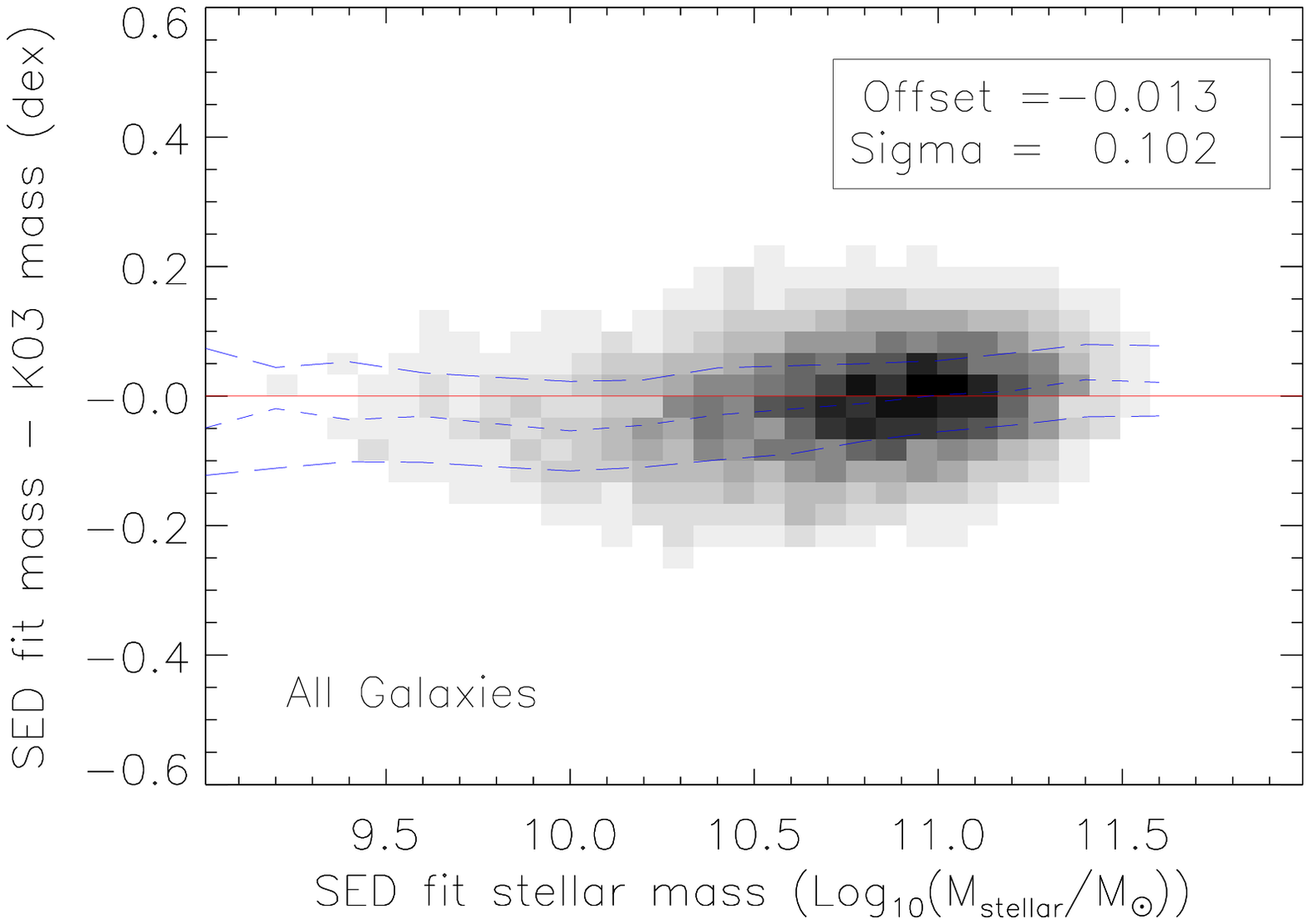}
\caption[]{The difference between the SED fit masses
  presented in this paper and those based on
  \citet[K03]{kauffmann_mass}. The masses are well reproduced using our
  method. The red line represent the equality of the masses, while the
  blue dashed lines show the median and 1 sigma upper and lower bounds
  of the distribution.}
\label{fig-mpamymass}
\end{figure}

In Figure \ref{fig-mpamymass}, we show the difference between our
stellar masses and those of K03 as a function of our stellar mass. The
K03 masses have been converted from a \citet{kroupa} to a Chabrier IMF
by dividing by 1.04.  The masses are well reproduced using our method,
with a scatter that is smaller than the average of the uncertainties
in either method. This likely means that despite our different
methods, the uncertainties on each method are correlated. Nonetheless,
our SED fitting seems to produce reliable results, notwithstanding
systematic issues common in both methods (ie. evolving IMF, stellar
population synthesis uncertainties, etc.).

\subsection{SDSS comparisons: Star formation rates}
\begin{figure*}
\includegraphics[width=\textwidth]{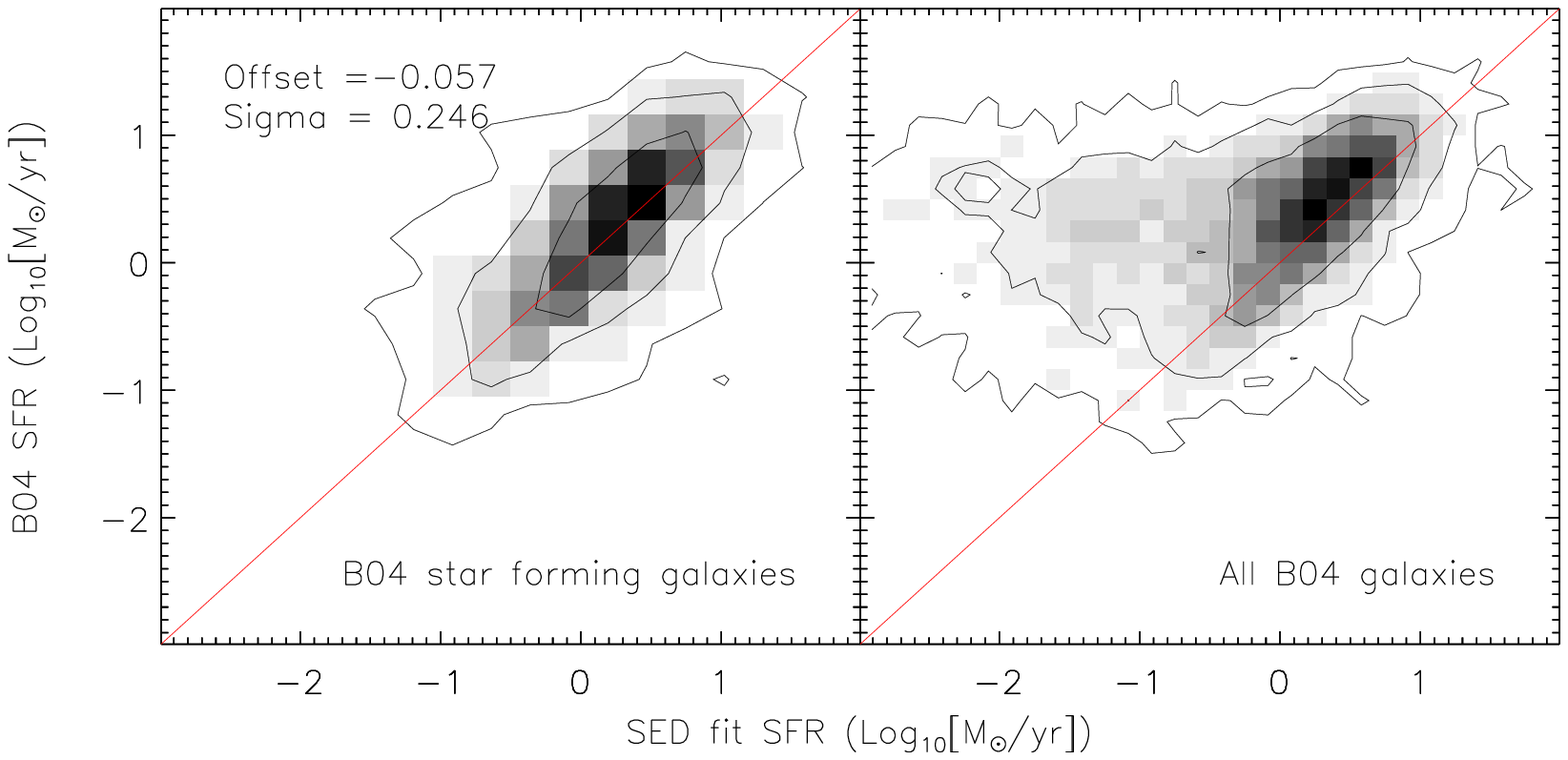}
\caption[]{A comparison between the SED fit
  star formation rates in this paper and those found by
  \citet[B04]{brinchmann_sfr}. We show this for galaxies that B04
  classifies as ``star-forming'' in the left panel and all galaxies in
  the right panel. The red line shows the equality between the B04 and
  SED fit star formation rates.}
\label{fig-mpamysfr-all}
\end{figure*}

One of the most common and robust methods of determining star
formation rates of star forming galaxies is to use the recombination
emission line H$\alpha$, which re-radiates the radiation of massive
young stars. To that end, B04 use the emission lines of SDSS galaxies,
to both classify the type of galaxy and to determine the star
formation rates. As this method is completely based on spectral
features this should give essentially an independent measurement to
ours.

B04 classifies galaxies into several categories based on the position
of the galaxy within the \citet[BPT]{Baldwin+81} diagram and the
quality of the spectra. The BPT diagram is used to determine whether a
galaxy has AGN activity based on the ratios of four emission lines
([OIII]5007, H$\beta$, [NII]6584 and H$\alpha$). A galaxy with a high
signal-to-noise ratio in each emission line and that lies in the star
forming locus of the BPT diagram is defined as a ``star forming''
galaxy. The other B04 categories are used to denote AGN or Composite
galaxies, as well as low S/N categories of each.

The star formation rate of high signal-to-noise star forming galaxies
is relatively straightforward to obtain from emission lines. Thus, it
is this class of galaxies that we first compare our SED fit star
formation rates to, as shown in the left panel of Figure
\ref{fig-mpamysfr-all}. The B04 star formation rates were calculated
assuming a \citet{kroupa} initial mass function. Following
\citet{Salim+07}, we have divided the B04 rates by 1.06 to account for
this difference in IMF. Our measurement agrees very well, with only a
negligible offset and a 1 $\sigma$ value that is consistent with our
expected measurement errors. Importantly, some of this scatter is
likely real, given that the B04 SFRs are based on H$\alpha$, which is
sensitive to star formation within the last 10 Myrs, while our SFRs
are based on $NUV$, which is sensitive star formation in the last 100
Myrs. An additional source of possible scatter arises because the SDSS
survey spectra, which are made by a fiber based spectrograph, only
sample the central 1/3 of the total galaxy light at the median
redshift. Although B04 attempts to correct these aperture effects by
using the available photometry, these will likely add further scatter
when compared to our method of using total magnitudes alone.

The right panel of Figure \ref{fig-mpamysfr-all} shows the comparison
of our SFRs with all of the galaxies for which B04 determined SFRs and
are in the \galex sample. In other words, in addition to those
galaxies B04 classified as star forming, this panel also includes
galaxies classified as AGN, composite or low S/N. While the SED fit
high star formation rates still agree well with the B04 measurements,
there is a long tail of SED fit determined low star formation rates
that in B04 are forming stars at much higher rates. \citet{Salim+07}
also noted this behavior when comparing UV derived SFR with B04. Salim
\etal extensively investigate this phenomenon and find that these
galaxies are ones in which B04 used indirect SF indicators, either
because of lack of H$\alpha$ emission or because of AGN contamination
of the nebular emission lines. Thus, as Salim \etal conclude, the UV
SED fit SFR are likely more accurate for these galaxies. Further, the
indirect methods used by B04 (ie. D$_n$(4000)-based) measure SFRs over
different timescales than the Balmer lines, and our UV-based method.

\subsection{ GEEC comparisons: Stellar Masses  }

In \citet{Balogh_smass}, we determined stellar masses for a limited
sample of GEEC galaxies based on the $K$ band data that was available
at the time. These were determined by computing $K$ band mass to light
ratios using simple stellar population modeling. The authors use a
young, constantly star forming galaxy model to approximate the most
blue galaxies ($B$ - $V$ $<$ 0.4) and a old, single stellar population
model for the reddest galaxies ($B$ - $V$ $>$ 1). The mass to light
ratios for intermediate galaxies are determined by a linear
interpolation between these two extremes based on the galaxy's $B$-$V$
color.

\begin{figure}
\includegraphics[width=8cm]{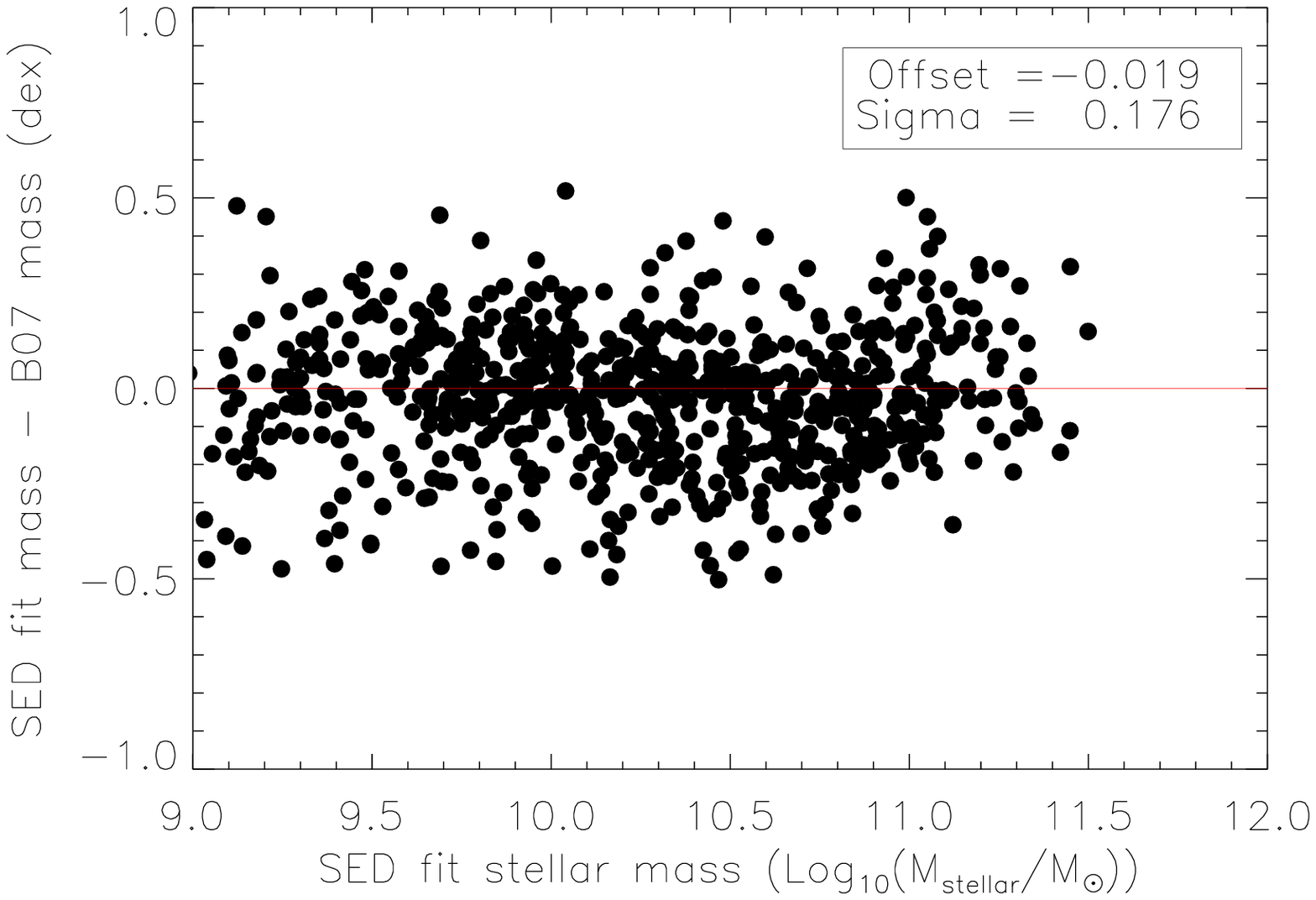}
\caption[Comparison between SED fit stellar masses and those
  determined from $K$-band data in \citet{Balogh_smass}]{Comparison
  between the SED fit stellar masses in this paper and the $K$-band
  determined \citet{Balogh_smass}(B07) masses for galaxies in both
  catalogs in the redshift range 0.1 $<$ z $<$ 0.55. The 1 $\sigma$
  scatter is 0.176 dex, and the SED fit stellar masses are
  systematically lower by 0.019 dex (ie. $\sim$ 2 $\%$). }
\label{fig-b07massmine}
\end{figure}

In Figure \ref{fig-b07massmine}, we present a comparison between our
current SED fit stellar masses and those of B07. We have converted
from a Salpeter IMF to a Chabrier IMF by subtracting 0.21
dex. Encouragingly, there is only a slight offset of -0.019 between
the masses and the scatter is quite small 0.176 $\sigma$.  This
scatter is actually smaller than the formal errors of both mass
measurements, but since they are, at least partially, using the same
photometry, the errors are correlated.

\end{document}